\documentclass[english,a4paper]{article}
\usepackage{dcolumn}
\usepackage{bm}
\usepackage{graphicx}
\usepackage{amsmath}
\usepackage{amsfonts}
\usepackage{amssymb}
\usepackage{amsthm}
\usepackage{amstext}
\usepackage{amsbsy}
\usepackage{amsopn}
\usepackage{amscd}
\usepackage{amsxtra}

\def\Id{\textrm 1\!\!\!\!1}
\def\openone{\leavevmode\hbox{\small1\kern-3.3pt\normalsize1}}
\DeclareMathOperator*{\am}{am}
\DeclareMathOperator*{\K}{K}
\DeclareMathOperator*{\Fell}{F}

\DeclareMathOperator*{\sign}{sign}
\usepackage[usenames,dvipsnames]{color}

\begin{document}
\title{Robust optimal control of two-level quantum systems}
\author{L. Van Damme, Q. Ansel, S. J. Glaser\footnote{Department of Chemistry, Technische Universit\"at
M\"unchen, Lichtenbergstrasse 4, D-85747 Garching, Germany}, D. Sugny\footnote{Laboratoire Interdisciplinaire Carnot de
Bourgogne (ICB), UMR 6303 CNRS-Universit\'e Bourgogne-Franche Comt\'e, 9 Av. A.
Savary, BP 47 870, F-21078 Dijon Cedex, France and Institute for Advanced Study, Technische Universit\"at M\"unchen, Lichtenbergstrasse 2 a, D-85748 Garching, Germany, dominique.sugny@u-bourgogne.fr}}

\maketitle

%
\begin{abstract}
We investigate the time and the energy minimum optimal solutions for the robust control of two-level quantum
systems against offset or control field uncertainties. Using the Pontryagin Maximum Principle, we derive the global optimal pulses for the first robustness orders. We show that the dimension of the control landscape is lower or equal to $2N$ for a field robust to the $N$th order, which leads to an estimate of its complexity. 
\end{abstract}


\section{Introduction}
Quantum control techniques are nowadays at the core of emergent quantum technologies in a multitude of domains extending from molecular and solid state physics to Nuclear Magnetic Resonance (NMR) and Magnetic Resonance Imaging \cite{glaserreview,brifreview,altafinireview,dongreview,calarco}. One of the main obstructions to the experimental realization of open-loop control processes is their high sensitivity to experimental imperfections and model
uncertainties. Since the start of quantum control, this question has motivated the development of pulse design methods addressing such robustness issues \cite{brumerbook,ricebook,alessandrobook}. Adiabatic quantum control techniques were first applied with success in some examples, but these protocols have a limited efficiency in a general setting due to the requirement of high energy and long duration fields \cite{adiabaticreview,adiabaticrmp}. Composite pulses \cite{compvitanov,complevitt,composite} and shortcut to adiabaticity techniques \cite{STA,STAnjp,daemsprl} have also been proposed, but they cannot reach the physical limits of the dynamical process in terms of time or efficiency. We will show that Optimal Control Theory (OCT) can be a perfect tool to overcome these difficulties \cite{alessandrobook,pont,bonnardbook,jurdjevicbook}. OCT is a general approach allowing to manipulate the system dynamics by determining the control field that minimizes a cost functional which can be, e.g., the control duration or its energy \cite{grape,reichkrotov,gross}. However, optimal control fields are not robust by construction and this issue is still at the center of a vivid debate \cite{glaserreview,brifreview,rabitz90,ticozzi}. Different numerical approaches ranging from the simultaneous control of an inhomogeneous ensemble of quantum systems \cite{kobzar1,kobzar2,kobzar3,rabitzrobust,rabitzrobust2,khanejaens} to pseudo-spectral methods \cite{li1,li2} have been proposed. Only local optimal solutions are obtained, with no certitude about the global optimality of the control process. In this work, we show how this fundamental question can be solved through the Pontryagin Maximum Principle (PMP) \cite{pont,bonnardbook,jurdjevicbook}. This approach has already been used with success in different optimal quantum problems \cite{alessandro,boscain,lapertprl,garon,khanejaspin,contrast,mesure,stefanatos,magicplane,zhang} and we propose here to extend its range of application to the design of robust control protocols.
The PMP transforms the optimal control problem into a generalized Hamiltonian system subject to a maximization condition and some boundary constraints. In this framework, the goal consists in finding the Hamiltonian trajectory reaching the target state, while minimizing the cost functional which defines the optimization procedure. A key advantage of the PMP is that it reduces the initial infinite-dimensional control landscape \cite{brifreview,landscape} to a finite space of low dimension. As shown below, this property is crucial in the search for globally optimal controls.

In this paper, we establish time and energy minimum optimal control strategies leading to a robust and precise state to state transfer of two-level quantum systems. The implementation of quantum gates is also analyzed. The measure of the robustness is given by the deviation of the control fidelity against offset or field inhomogeneities. This description reproduces the standard experimental uncertainties that can be encountered in quantum information processing, in NMR and in atomic or molecular physics~\cite{glaserreview,spin,ernst,chuang}. The robustness is defined either locally by expanding the state of the system order by order with respect to the unknown parameters~\cite{STAnjp,daemsprl} or globally by considering a discrete inhomogeneous ensemble of quantum systems~\cite{kobzar1,kobzar2}. A precise definition will be given later. Ultra-precise or broadband excitation profiles are respectively realized based on the first or second measure, which will be called local or broadband robustness below. 
The two definitions will be considered in the different examples. Note that the controllability of the different systems is assumed and not discussed in this work~\cite{khaneja1,khaneja2}.

The paper is organized as follows. Section~\ref{sec2} introduces the model we study. The optimal solutions for the energy and time-minimum inversion robust against offset inhomogeneities are presented in Sec.~\ref{sec3}. Section~\ref{sec4} focuses on the robustness with respect to control field imperfections, while Sec.~\ref{sec5} is dedicated to the broadband control of an ensemble of spins. A comparison with the results obtained with a numerical optimization algorithm is made in Sec.~\ref{sec6}. The method is generalized in Sec.~\ref{SectGate} to the robust implementation of one-qubit gates. Conclusion and prospective views are given in Sec.~\ref{conc}. Technical computations are reported in the Appendices \ref{appa}, \ref{appb} and \ref{appc}.
\section{The model system}\label{sec2}
We consider the Bloch representation of a two-level quantum system whose dynamics is governed by the Bloch equation. The Bloch vector $\vec{q}(t)={^t(x,y,z)}$ satisfies the following differential system:
\begin{equation}\label{eq1}
\dot{\vec{q}}(t)=\begin{pmatrix}
0 & \delta & -(1+\alpha)u_y\\ -\delta & 0 & (1+\alpha)u_x\\ (1+\alpha)u_y & -(1+\alpha)u_x & 0
\end{pmatrix}\vec{q}(t),
\end{equation}
where $u_x$ and $u_y$ are the two control fields. The parameters $\delta$ and $\alpha$ represent respectively the offset and control field inhomogeneities. We first consider the case where $\alpha=0$. We assume that the solution of the Bloch equation can be written as a perturbative expansion in $\delta$ up to a given order:
\begin{equation}\label{eq2}
\vec{q}(t)=\vec{q}_0(t)+\delta\vec{q}_1(t)+\cdots +\delta^N\vec{q}_N(t)+O(\delta^{N+1}),
\end{equation}
with $\vec{q}_i={^t(x_i,y_i,z_i)}$. The vector $\vec{q}_0$ is the homogeneous part of the solution and $\vec{q}_i$ the inhomogeneous contribution due to the $i$th- order term of the expansion. We investigate the robust control of the inversion of the Bloch vector, i.e. the goal is to bring in a time $t_f$ the state $\vec{q}(t)$ from the north pole to the south pole of the Bloch sphere. Other initial or target states can be analyzed in the same way. The control process can be expressed in the perturbative expansion as:
\begin{equation}
\begin{aligned}
& \vec{q}_0(0)={^t(0,0,1)} \mapsto \vec{q}_0(t_f)={^t(0,0,-1)},\\
& \vec{q}_{i}(0)={^t(0,0,0)} \mapsto \vec{q}_{i}(t_f)={^t(0,0,0)},~\textrm{for}~i\in\{1,\cdots,N\}.
\end{aligned}
\label{TransInvRob}
\end{equation}
Note that the target states of the inhomogeneous contributions ensure that the offset term does not modify the final state of the system up to the order $N$ in $\delta$, improving thus the robustness of the control protocol. Plugging Eq.~\eqref{eq2} into Eq.~\eqref{eq1}, it is straightforward to show that the differential system governing the dynamics of each vector $\vec{q}_i$ is given by:
\begin{equation}
\frac{d}{dt}\begin{pmatrix}
\vec{q}_0\\ \vec{q}_1\\ \vec{q}_2 \\ \vdots \\ \vec{q}_N
\end{pmatrix}=\begin{pmatrix}
H_0 & 0 & 0 & \cdots & 0\\
\partial_{\delta} H & H_0 & 0 & & 0\\
0 & \partial_{\delta}H & H_0 & & 0\\
\vdots & & \ddots & \ddots & \vdots\\
0 & \cdots & 0 & \partial_{\delta}H & H_0
\end{pmatrix}\begin{pmatrix}
\vec{q}_0\\ \vec{q}_1\\ \vec{q}_2 \\ \vdots \\ \vec{q}_N
\end{pmatrix},
\label{Dsystdelta}
\end{equation}
where:
\begin{equation}
H_0=\begin{pmatrix}
0 & 0 & -u_y\\ 0 & 0 & u_x\\ u_y & -u_x & 0
\end{pmatrix},\quad\partial_{\delta}H=\begin{pmatrix}
0 & 1 & 0\\ -1 & 0 & 0\\ 0 & 0 & 0
\end{pmatrix}.
\end{equation}
The originality of the method consists in directly solving the PMP applied to Eq.~\eqref{Dsystdelta}. We show that the optimal control fields must satisfy a certain differential system depending upon a finite number of parameters for a robustness at order $N$. This number can be interpreted as the dimension of the control landscape. A careful investigation of this landscape allows us to detect the global optimal solution of the process, at least for low robustness orders.
\section{Energy and time minimum inversion of two-level quantum systems robust against offset inhomogeneities}\label{sec3}
We consider in this paragraph the robust optimal inversion with respect to offset uncertainties. In this case, the PMP is formulated from the pseudo-Hamiltonian $H_P$ which can be written as follows \cite{pont,bonnardbook,jurdjevicbook}:
\begin{equation}
H_P=\sum_{i=0}^N \vec{p}_i\cdot \dot{\vec{q}}_i+p^0f^0.
\end{equation}
This leads to:
\begin{equation}\label{pseudo}
H_P=\vec{p}_0\cdot (\vec{q}_0\times \vec{u})+\sum_{k=1}^n \vec{p}_k\cdot (\vec{q}_k\times \vec{u}+
\vec{q}_{k-1}\times \vec{e}_z)+p^0 f^0,
\end{equation}
where $\vec{p}_i$ is the adjoint state of $\vec{q}_i$ and $p_0$ a negative constant, which is set to $-1/2$ and to $-1$ for the energy and time optimal control problems, respectively~\cite{pont,bonnardbook,jurdjevicbook}. $f^0$ is a function of $u_x$ and $u_y$ whose integral over time gives the associated cost functional $C$ to minimize. We have:
$$
C_E=\int_0^{t_f}f^0(u_x,u_y)dt=\int_0^{t_f}[u_x^2+u_y^2]dt,
$$
for the energy and
$$
C_t=\int_0^{t_f}dt=t_f
$$
for the time, where the control duration $t_f$ is not fixed. In Eq.~\eqref{pseudo}, the vectors $\vec{u}$ and $\vec{e}_z$ have the coordinates $(u_x,u_y,0)$ and $(0,0,1)$ and $\times$ denotes the vector product of two three-dimensional vectors.

We introduce the angular momenta $\vec{\ell}_{a,b}=\vec{p}_a\times \vec{q}_b$ and the partial sums $\vec{\Omega}_n=(\Omega_{nx},\Omega_{ny},\Omega_{nz})=\sum_{i=k}^n\vec{\ell}_{i,i-k}$. Using the properties of the scalar triple product, we arrive at:
$$
H_P=\vec{u}\cdot \vec{\Omega}_0+\vec{e}_z\cdot \vec{\Omega}_1+p^0f^0.
$$
The PMP states that the coordinates of the Bloch vector $\vec{q}$ and of the corresponding adjoint state $\vec{p}$ fulfill the Hamiltonian's equations associated with $H_P$:
$$
\dot{\vec{q}}=\frac{\partial H_P}{\partial \vec{p}},~\dot{\vec{p}}=-\frac{\partial H_P}{\partial \vec{q}},
$$
the control fields being given by the maximization condition~\cite{pont,bonnardbook,jurdjevicbook}:
$$
H(\vec{x},\vec{p})=\max_{(u_x,u_y)\in U}H_P(\vec{q},\vec{p},u_x,u_y).
$$
The set $U$, which defines the constraint on the pulses, is given by $U=\mathbb{R}^2$ and by $u_x^2+u_y^2\leq 1$ for the energy and time minimization problems, respectively. Note that the constraint $u_x^2+u_y^2\leq 1$ avoids the occurrence of very intense control fields which are not relevant experimentally. We obtain $H=\frac{\Omega_{0x}^2+\Omega_{0y}^2}{2}+\Omega_{1z}$, with $u_x=\Omega_{0x}$ and $u_y=\Omega_{0y}$, for the energy-minimum and $H=r+\Omega_{1z}$, with $u_x=\Omega_{0x}/r$, $u_y=\Omega_{0y}/r$ and $r=\sqrt{\Omega_{0x}^2+\Omega_{0y}^2}$, for the time-minimum problem.

Using the maximization condition of the PMP, a straightforward computation then leads to:
\begin{equation}
\begin{cases}
\dot{\vec{\Omega}}_0=\vec{\Omega}_{1}\times\vec{e}_z,\\
\dot{\vec{\Omega}}_k=\vec{\Omega}_k\times\vec{\Omega}_0+ \vec{\Omega}_{k+1}\times\vec{e}_z,~k\in\{1,\cdots,N-1\}\\
\dot{\vec{\Omega}}_N=\vec{\Omega}_N\times\vec{\Omega}_0,
\end{cases}
\label{edEopt}
\end{equation}
in the energy case and to
\begin{equation}
\begin{cases}
\dot{\vec{\Omega}}_0=\vec{\Omega}_{1}\times\vec{e}_z,\\
\dot{\vec{\Omega}}_k=\frac{1}{r}\vec{\Omega}_k\times\vec{\Omega}_0+ \vec{\Omega}_{k+1}\times\vec{e}_z,~~k\in\{1,\cdots,N-1\}\\
\dot{\vec{\Omega}}_N=\frac{1}{r}\vec{\Omega}_N\times\vec{\Omega}_0,
\end{cases}
\label{edTopt}
\end{equation}
for the time-minimum problem. In the two situations, we deduce that $\Omega_{0z}(t)=cst$. This constraint becomes $\Omega_{0z}(t)=0$ when the initial point is the north pole of the Bloch sphere. The differential systems \eqref{edEopt} and \eqref{edTopt} can be interpreted as the conditions to satisfy by the control fields to realize the control process.
Moreover, the initial phase of the control fields is irrelevant, which means that we can set $\Omega_{0y}(t=0)=0$. Another point is related to the fact that the norm $|\vec{\Omega}_N|$ is constant in time, and can be set to $1$ without loss of generality (this is equivalent to a time rescaling). We obtain that $\vec{\Omega}_N(t=0)$ only depends on one angle, i.e. $\vec{\Omega}_N(0)=(\cos\vartheta,\sin\vartheta,0)$. Finally, the field depends on $2N$ parameters at order $N$: $\Omega_{0x}(0)$, $\Omega_{kx}(0)$ and $\Omega_{ky}(0)$ for $k\in\{1,\cdots N-1\}$, and the angle $\vartheta$. This number of parameters is also the dimension of the control landscape. The last point to solve is to adjust these parameters to realize the transfer~\eqref{TransInvRob}. Many different solutions can exist. For $N$ small enough, the low dimension of the control landscape allows us to find the global optimal solution minimizing $C_E$ or $C_t$.

We now describe the results for the energy-minimum problem. We consider here only one control field along $\vec{e}_x=(1,0,0)$, i.e. $u_y(t)=0$. We compute the global optimal solutions for the first, second and third robustness orders. The analytical expression of the control field is derived at first order, while numerical optimization techniques are used for the second and third order. The computations are detailed in the Appendix~\ref{appa}. Figure~\ref{figEopt} displays the control fields and the homogeneous contribution $\vec{q}_0$ of the Bloch vector.
\begin{figure}[h!]
\centering
\includegraphics[angle=-90,scale=0.4]{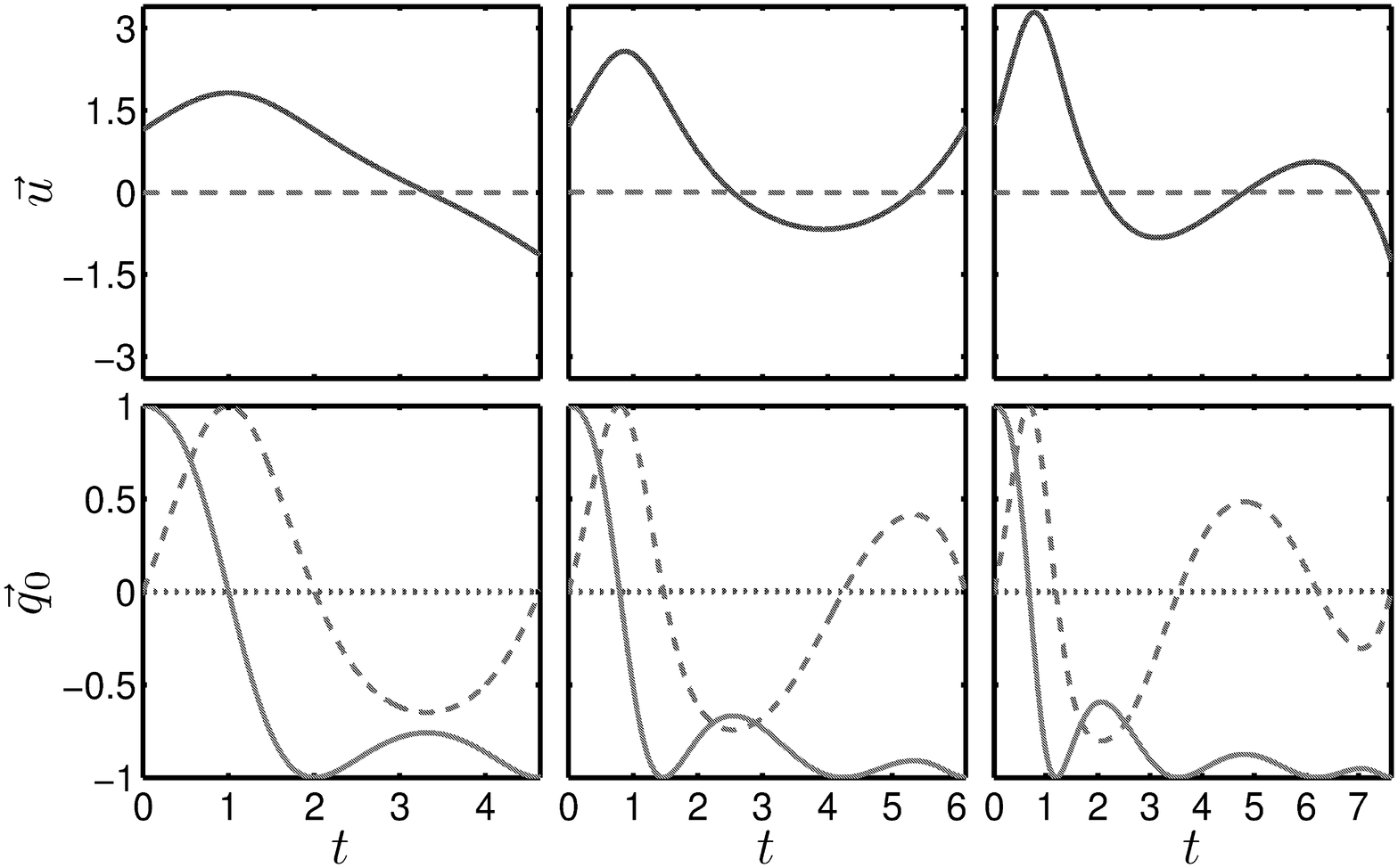}
\caption{\textit{Upper panels:} Control fields of minimum energy robust to first order (left), to second order (middle), and to third order (right) in $\delta$. The solid and dashed lines depict respectively the fields $u_x(t)$ and $u_y$. \textit{Lower panels:} Evolution of the components of the homogeneous solution $\vec{q}_0={^t(x_0,y_0,z_0)}$ (which corresponds to the on-resonance case). The dotted, dashed and solid lines represent respectively the dynamics of $x_0(t)$, $y_0(t)$ and $z_0(t)$.\label{figEopt}}
\end{figure}
We observe the peculiar dynamics of the homogeneous part of the Bloch vector, which oscillates in the $(y_0,z_0)$ plane like a damped oscillator, with a number of oscillations increasing with the robustness order in $\delta$. The efficiency of the control protocol is shown in Fig.~\ref{figrobEopt}. As could be expected, a better robustness is achieved when higher orders are nullified.
\begin{figure}[h!]
\centering
\includegraphics[angle=-90,scale=0.4]{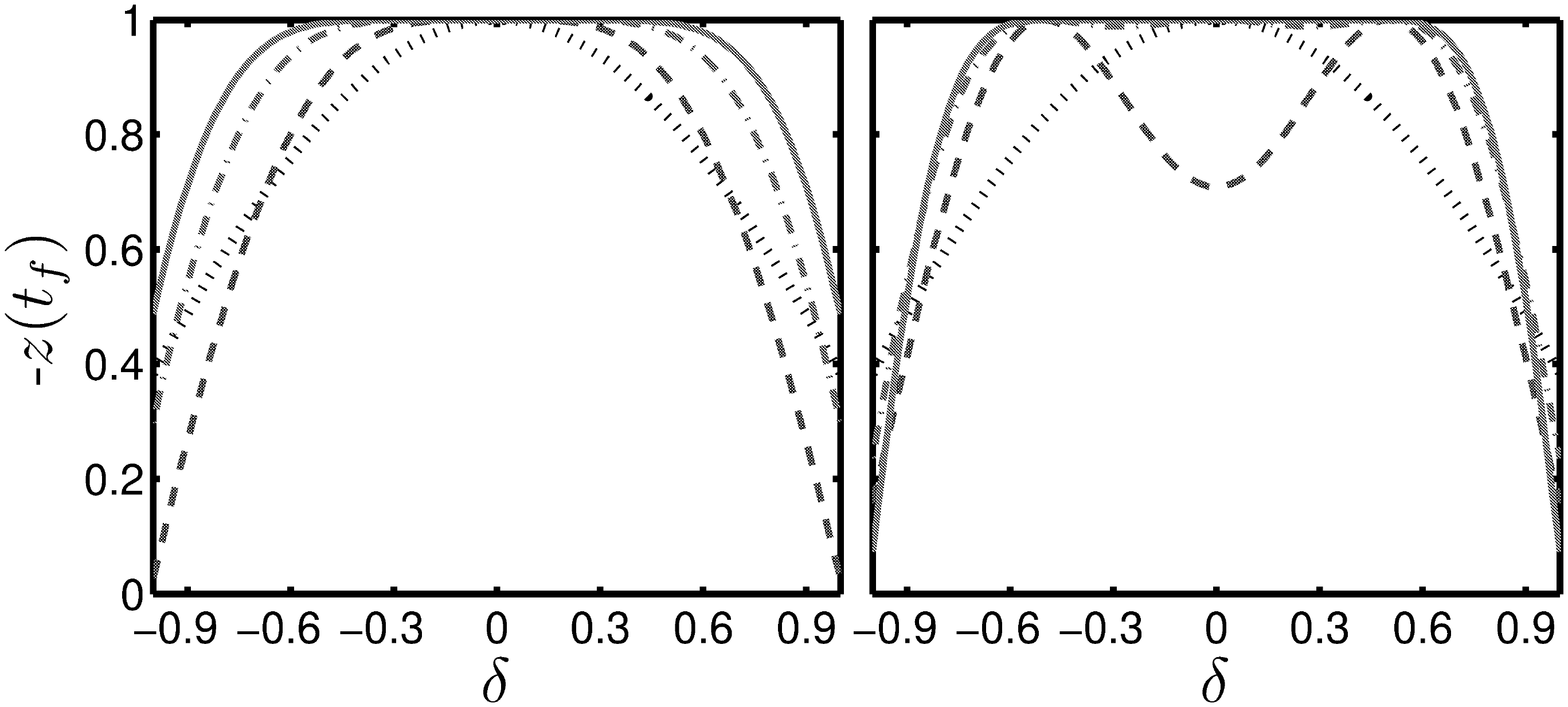}
\caption{(left panel) Fidelity $F=-z(t_f)$ for the local robustness of the inversion transfer as a function of the offset $\delta$ for a standard $\pi$-pulse (dotted line), the first-order optimal robust solution (dashed line), the second order one (dashed-dotted line), and the third order one (solid line). (right panel) Same as the left panel but for the broadband robustness.  The cases of two-, three- and four two-level quantum systems are displayed respectively in dashed, dashed-dotted and solid lines. The offsets are set to $\pm 0.5$, $\pm 0.5$ and 0, $\pm 1/6$ and $\pm 0.5$ for the three examples.\label{figrobEopt}}
\end{figure}

In the time-optimal case, it can be shown that the minimum time to cancel the first robustness order is associated with a pulse of constant intensity along the $x$- direction of the Bloch sphere. This pulse switches between the values 1 and -1 at time $t=3\pi/2$ and it has a total duration of $t_f=2\pi$. The pulse structure is not the same at second and third orders. The robust inversion is realized by smooth fields of durations $t_f=2.44\pi$ and $t_f=3.54\pi$ for the second and third order robust control fields. The details of the computations are given in Appendix~\ref{appb}. Figure~\ref{figTopt} displays the fields $u_x$ and $u_y$ and the homogeneous contribution $\vec{q}_0(t)$ of the Bloch vector. We recover with this geometrical analysis the transition found numerically in \cite{kobzar1,kobzar2} from a square signal to a smooth field. 
\begin{figure}[h!]
\centering
\includegraphics[angle=-90,scale=0.4]{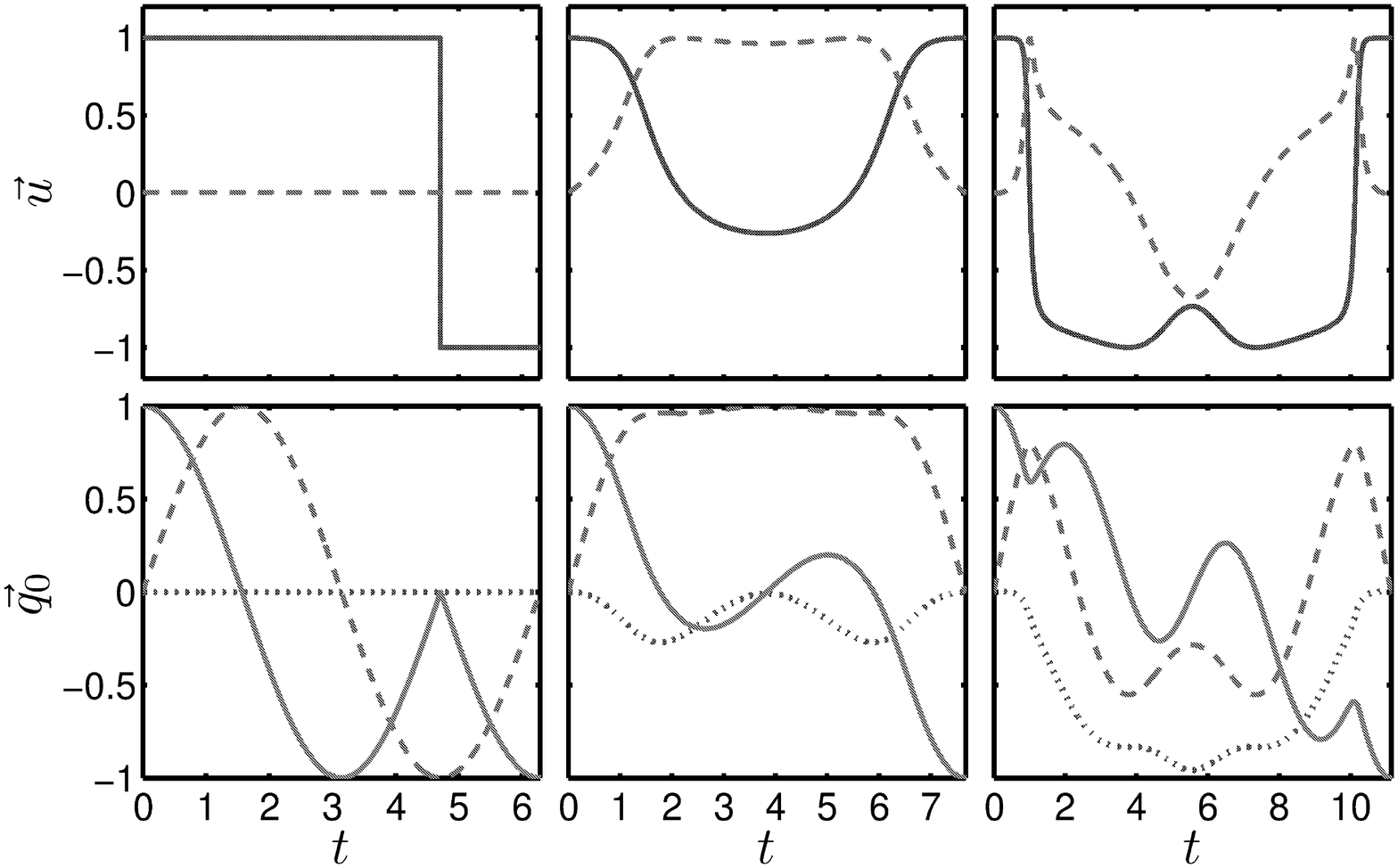}
\caption{Same as Fig. 1 but for the time-optimal solution. \label{figTopt}}
\end{figure}
The different numerical results are collected in Tab.~I.
\begin{table}[h!]\label{Tab1}
\centering
\begin{tabular}{|c|c|c|c|}
\hline\hline
Type & Cost
& Orders & Area or time ($\times \pi$)\\
\hline
$\delta$ & time & 1/2/3 & 2.0/2.44/3.54 \\
$\alpha$ & time & 1/2/3 & 1.86/2.71/3.56 \\
$\delta$ & energy & 1/2/3 & 1.45/1.81/2.11 \\
\hline\hline
\end{tabular}
\caption{Robustness of order $n$ with respect to the offset terms (type $\delta$) or to the control field inhomogeneities (type $\alpha$) for the local measure. The pulse area and the minimum time are given for the energy and time minimum control problems, respectively.}
\end{table}
\section{Robustness against control field inhomogeneities}\label{sec4}
The same method can be applied to the field inhomogeneities and the $\alpha$- parameter. We consider the on-resonance case where $\delta=0$. We assume that the solution can be written as a perturbative expansion of the form:
\begin{equation}
\vec{q}(t)=\vec{q}_0(t)+\alpha\vec{q}_1(t)+\cdots +\alpha^N\vec{q}_N(t)+O(\alpha^{N+1}).
\end{equation}
We obtain the following differential system:
\begin{equation}
\frac{d}{dt}\begin{pmatrix}
\vec{q}_0\\ \vec{q}_1\\ \vec{q}_2 \\ \vdots \\ \vec{q}_N
\end{pmatrix}=\begin{pmatrix}
H_0 & 0 & 0 & \cdots & 0\\
H_0 & H_0 & 0 & & 0\\
0 & H_0 & H_0 & & 0\\
\vdots & & \ddots & \ddots & \vdots\\
0 & \cdots & 0 & H_0 & H_0
\end{pmatrix}\begin{pmatrix}
\vec{q}_0\\ \vec{q}_1\\ \vec{q}_2 \\ \vdots \\ \vec{q}_N
\end{pmatrix},
\label{edqialpha}
\end{equation}
with:
\begin{equation}
H_0=\begin{pmatrix}
0 & 0 & -u_y \\ 0 & 0 & u_x\\
u_y & -u_x & 0
\end{pmatrix}.
\end{equation}
The pseudo-Hamiltonian of this system is given by $H_p=\sum_{i=0}^N\vec{p}_i\cdot\dot{\vec{q}}_i+p^0f^0$, where $p^0$ and $f^0$ depend on the cost functional to minimize. Introducing the angular momenta $\vec{\ell}_{ij}=\vec{p}_i\times\vec{q}_i$, and the vectors
\begin{equation}
\begin{aligned}
&\vec{\Omega}_k=\sum_{i=k}^N\vec{\ell}_{i,i-k}+\sum_{i=k+1}^N\vec{\ell}_{i,i-k-1},\quad k\in\{1,\cdots,N-1\},\\ &\vec{\Omega}_N=\vec{\ell}_{N,0},
\end{aligned}
\end{equation}
we can show, using the Hamilton's equations, that:
\begin{equation}
\frac{d}{dt}\begin{pmatrix}
\vec{\Omega}_0\\ \vec{\Omega}_1\\ \vec{\Omega}_2 \\ \vdots \\ \vec{\Omega}_N
\end{pmatrix}=\begin{pmatrix}
H_0 & H_0 & 0 & \cdots & 0\\[-1.5ex]
0 & H_0 & H_0 & & \vdots\\[-1.5ex]
0 & 0 & H_0 & \ddots & 0\\[-1ex]
\vdots & &  & \ddots & H_0\\
0 & 0 & 0 & \cdots & H_0
\end{pmatrix}\begin{pmatrix}
\vec{\Omega}_0\\ \vec{\Omega}_1\\ \vec{\Omega}_2 \\ \vdots \\ \vec{\Omega}_N
\end{pmatrix}.
\end{equation}
Pontryagin's Hamiltonian can be written as follows:
\begin{equation}
H_P=\vec{u}\cdot\vec{\Omega}_0+p^0f^0,
\end{equation}
where $\vec{u}=(u_x,u_y,0)$. In this case, the PMP leads to the same condition both for the energy-minimum and time-minimum control problems. For the minimization of the time, the maximization condition gives $u_x=\frac{\Omega_{0x}}{\sqrt{\Omega_{0x}^2+\Omega_{0y}^2}}$ and $u_y=\frac{\Omega_{0y}}{\sqrt{\Omega_{0x}^2+\Omega_{0y}^2}}$. We deduce that the Pontryagin's Hamiltonian reads (after absorbing the constant $p^0f^0=-1$ in the definition of the pseudo-Hamiltonian):
\begin{equation}
\tilde{H}=\sqrt{\Omega_{0x}^2+\Omega_{0y}^2}.
\end{equation}
If $\tilde{H}\neq 0$, we can consider the Hamiltonian:
\begin{equation}
H=\Omega_{0x}^2+\Omega_{0y}^2,
\end{equation}
which is set to $H=1$ without loss of generality. The control fields are then given by $u_x=\Omega_{0x}$ and $u_y=\Omega_{0y}$. The application of the PMP in the energy minimum case leads to the same expression. The first consequence of the PMP is that $u_x^2+u_y^2=1$ for any time $t$. The differential system can be written as:
\begin{equation}
\begin{cases}
\dot{\vec{\Omega}}_0=(\vec{\Omega}_0+\vec{\Omega}_1)\times\vec{u},\\
\dot{\vec{\Omega}}_1=(\vec{\Omega}_1+\vec{\Omega}_2)\times\vec{u},\\
\dot{\vec{\Omega}}_2=(\vec{\Omega}_2+\vec{\Omega}_3)\times\vec{u},\\
\vdots \\
\dot{\vec{\Omega}}_N=\vec{\Omega}_N\times\vec{u}.
\end{cases}
\label{edOmegaalpha}
\end{equation}
At time $t=0$, we have $\Omega_{kz}(0)=0$ since the initial state of the Bloch vector is the north pole of the Bloch sphere. Moreover, the initial phase of the control field is irrelevant for the inversion transfer, leading to $u_y(0)=\Omega_{0y}(0)=0$ and $u_x(0)=\Omega_{0x}(0)=1$. The dimension of the control landscape is $2N$, and this space can be parameterized by $(\Omega_{kx}(0),\Omega_{ky}(0))$ for $k\in\{1,\cdots,2N\}$. The results for the orders 1, 2 and 3 are displayed in Fig.~\ref{figalphaopt}. The analytical expression of the first-order robust solution is given in the appendix~\ref{appc}.
\begin{figure}[h!]
\centering
\includegraphics[angle=-90,scale=0.4]{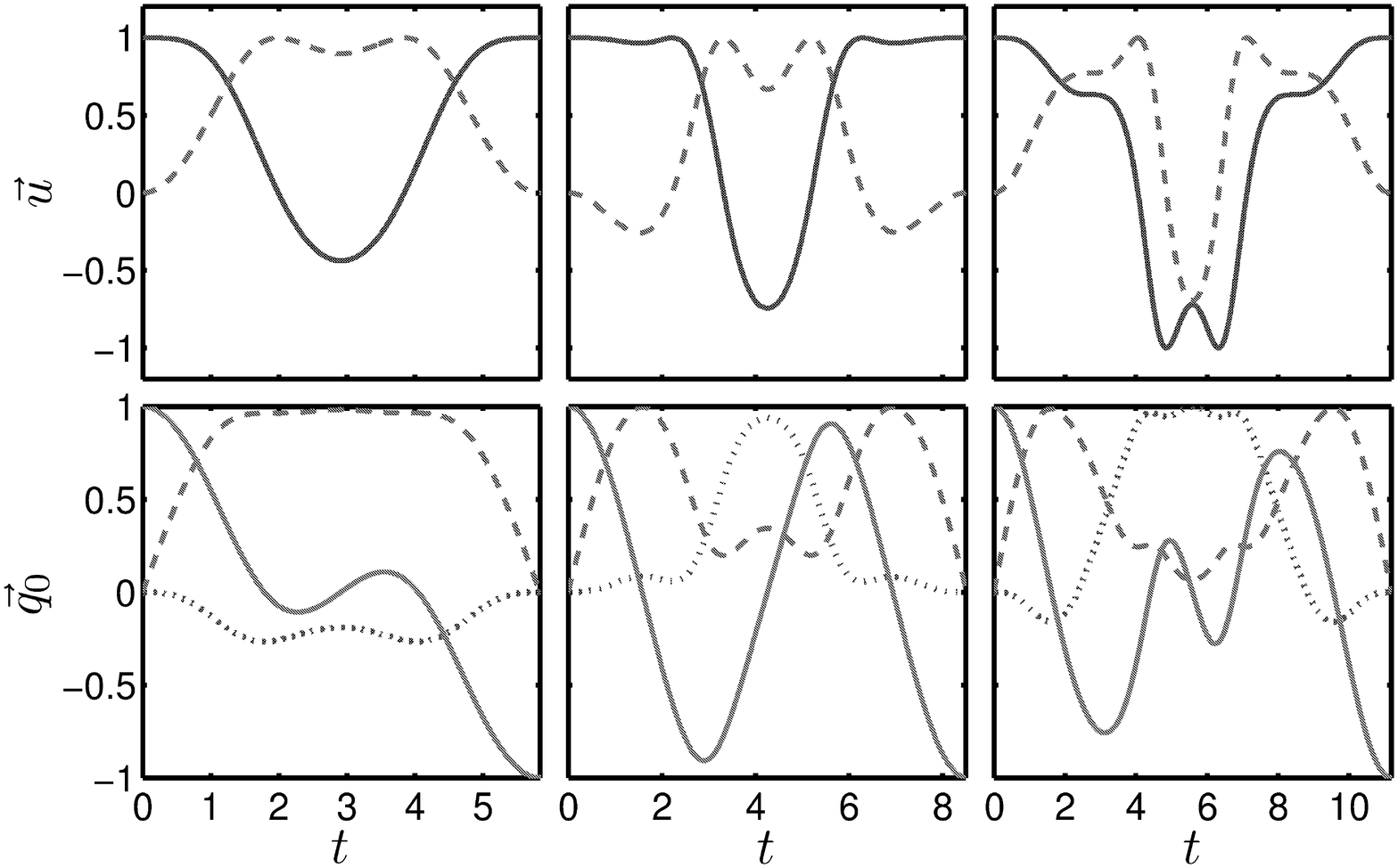}
\caption{Same as Fig. 1 but for the time-optimal solution robust against control field inhomogeneities.\label{figalphaopt}}
\end{figure}
The minimum times are given in Tab.~I. Note the linear evolution of these times as a function of the robustness order.

\section{Broadband robust optimal control}\label{sec5}
The robustness can also be defined through an inhomogeneous ensemble of quantum systems featured by different parameters. A finite number of systems belonging to this set is considered \cite{kobzar1,kobzar2,rabitzrobust,rabitzrobust2,khanejaens}. Our approach also works in this situation and leads to the same control landscape complexity. For an ensemble of quantum systems with different offsets, the dynamics of each element is of the form:
\begin{equation}
\dot{\vec{q}}_k(t)=\begin{pmatrix}
0 & \Delta_k & -u_y\\ -\Delta_k & 0 & u_x\\ u_y & -u_x & 0
\end{pmatrix}\vec{q}_k(t),
\label{edEnsembleopt}
\end{equation}
where $\Delta_k$ is the offset of the system $k$. We define the individual angular momentum $\vec{\ell}_k=\vec{p}_k\times\vec{q}_k$, where $\vec{p}_k$ is the adjoint state of the system $k$. The optimal control fields which invert simultaneously all the elements of the set satisfy:
\begin{equation}
\dot{\vec{\ell}}_k=\vec{\ell}_k\times\vec{u}+ \Delta_k\vec{\ell}_k\times\vec{e}_z,\quad k\in\{1,\cdots,N\},
\end{equation}
with $u_x=\sum_k \ell_{k,x}$, $u_y=\sum_k\ell_{k,y}$ for the energy-minimum problem, and $u_x=\frac{1}{r}\sum_k \ell_{k,x}$, $u_y=\frac{1}{r}\sum_k \ell_{k,y}$ with $r=\sqrt{\left(\sum_k \ell_{k,x}\right)^2+\left(\sum_k \ell_{k,y}\right)^2}$ for the time-minimum case. Since the initial point of each system is the north pole of the Bloch sphere, we deduce that $\ell_{k,z}=0$. A trajectory depends on $2N$ parameters but this number can be reduced to $2N-2$. Indeed, since the initial phase of the control is arbitrary, we can choose $u_y(0)=0$ which leads to $\sum_{i=1}^N\ell_{iy}(0)=0$. Moreover, one of the first integrals can be set to $1$ by rescaling the time. Here we choose $H=1$, which leads to $r(0)=\sum_{i=1}^N\ell_{ix}(0)=1$. Thus, we have $\ell_{Nx}(0)=1-\sum_{i=1}^{N-1}\ell_{ix}(0)$ and $\ell_{Ny}(0)=-\sum_{i=1}^{N-1}\ell_{iy}(0)$. Finally, we deduce that the total dimension of the control landscape is $2N-2$.

\begin{figure}[h!]
\centering
\includegraphics[angle=-90,scale=0.4]{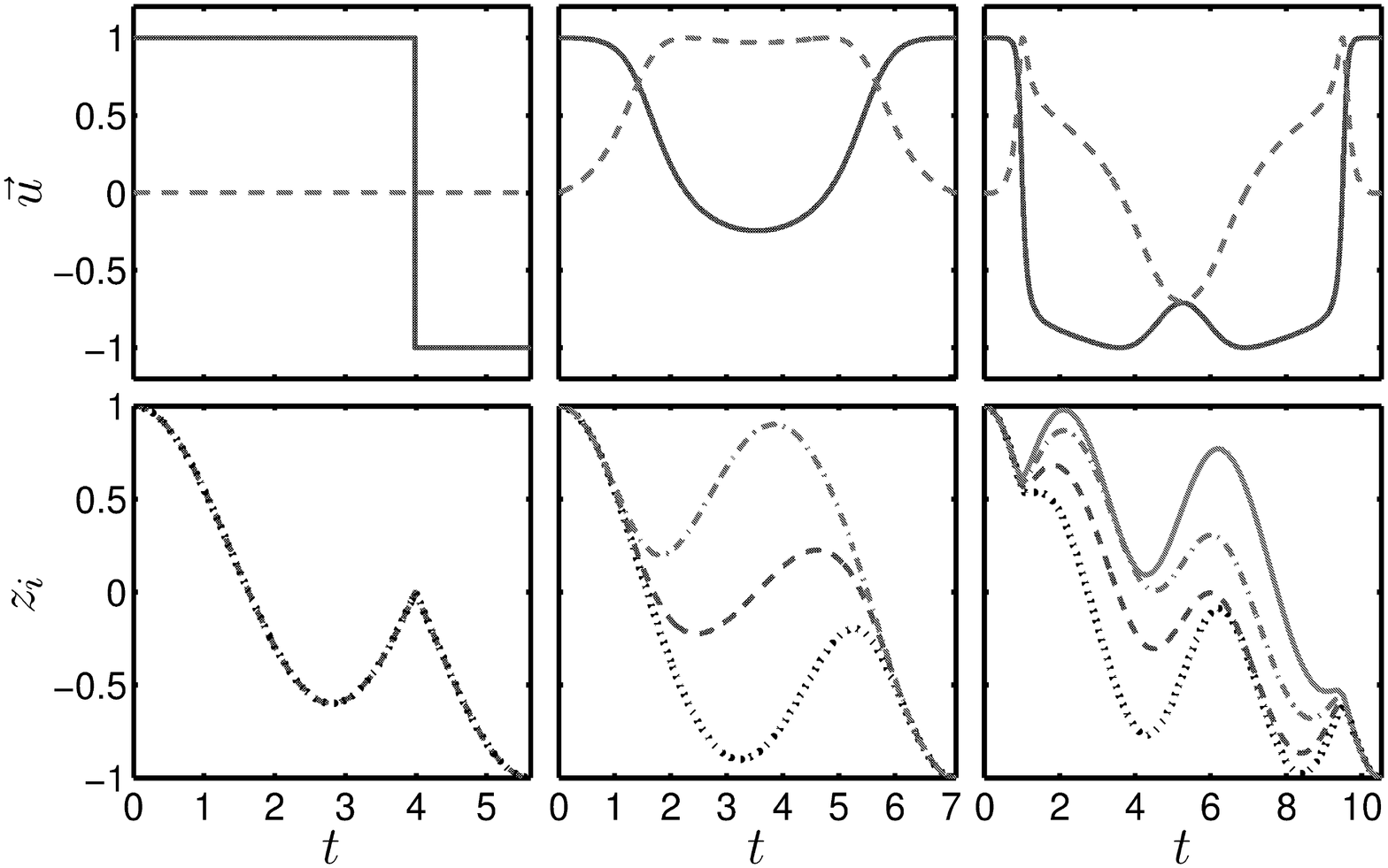}
\caption{Upper panels: Time-optimal control fields for the simultaneous control of two spin 1/2 particles with $\Delta_1=-0.5$ and $\Delta_2=0.5$, three spins with $\Delta_1=-0.5$, $\Delta_2=0$ and $\Delta_3=0.5$, and four spins with $\Delta_1=-0.5$, $\Delta_2=-1/6$, $\Delta_3=1/6$ and $\Delta_4=0.5$ (from left to right). Lower panels: Evolution of the $z$- coordinate of the corresponding Bloch vectors. $z_1(t)$ is associated with the black dotted line, $z_2(t)$ the dashed line, $z_3(t)$ the dash-dotted line and $z_4$ to the solid one.\label{fig5}}
\end{figure}

As displayed in Fig. \ref{fig5}, the time-optimal solutions for two, three and four quantum systems are very similar to the trajectories of Fig.~\ref{figTopt}.
For $N=2$, we recover the results derived in Ref.~\cite{assemat}.
The right panel of Fig.~\ref{figrobEopt} displays the inversion profile against the offset $\delta$. Note the very large robustness obtained although only four quantum systems are considered.
\section{Comparison with the GRAPE algorithm}\label{sec6}
The broadband inversion has been investigated numerically in Ref.~\cite{kobzar1,kobzar2}. The authors used the GRAPE algorithm~\cite{grape}, which is based on the PMP. However, the pulses were not optimized in the same way, in the sense that this algorithm aims at controlling a large number of spins ($\simeq$~100) over a certain range of offset inhomogeneities and for different pulse durations. Here, the goal is to compare this method to the time-optimal control of two, three and four spins. We recall that the control fields are so that $u_x=\cos\Phi(t)$ and $u_y=\sin\Phi(t)$. As a consequence, we use the GRAPE algorithm to optimize only the phase of the control fields (see Fig. 5 of Ref.~\cite{kobzar1}).

\begin{figure}[h!]
\centering
\includegraphics[angle=-90,scale=0.4]{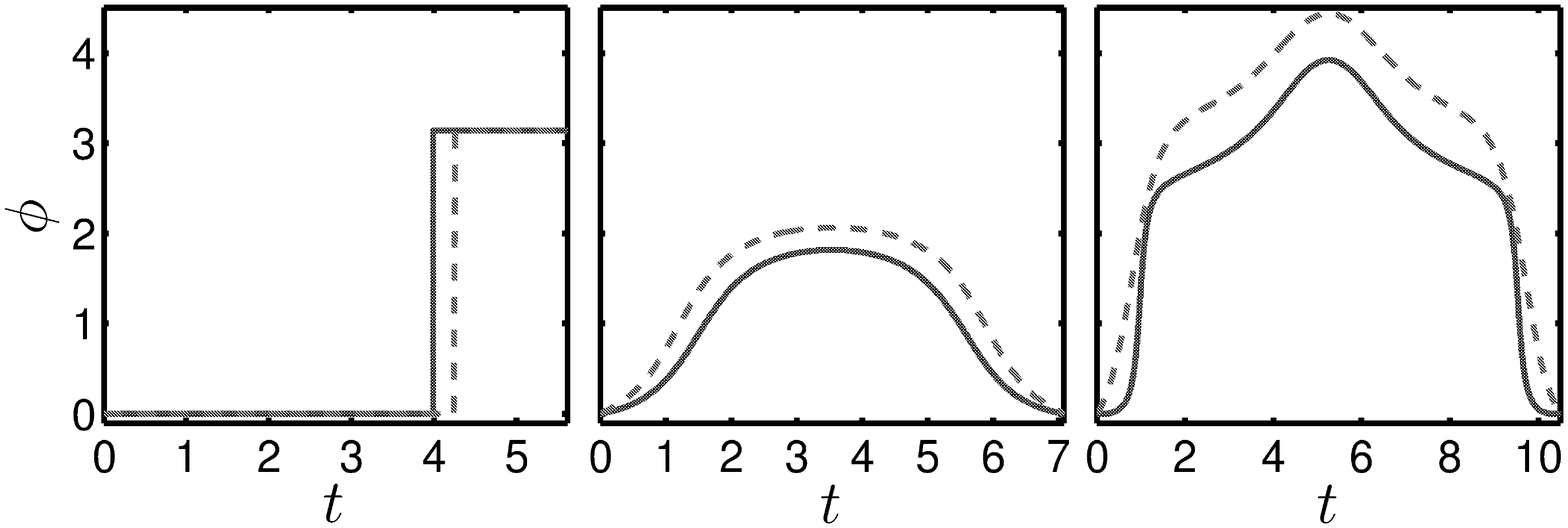}
\caption{\emph{From left to right:} Time-optimal phase of the control field for the inversion of two spins with $\Delta_i\in\{-1/2,1/2\}$, three spins with $\Delta_i\in\{-1/2,0,1/2\}$ and four spins with $\Delta_i\in\{-1/2,-1/6,1/6,1/2\}$ (solid line), and phase of the control field derived with a GRAPE optimization (dashed line).\label{figA9}}
\end{figure}

\begin{figure}[h!]
\centering
\includegraphics[angle=-90,scale=0.4]{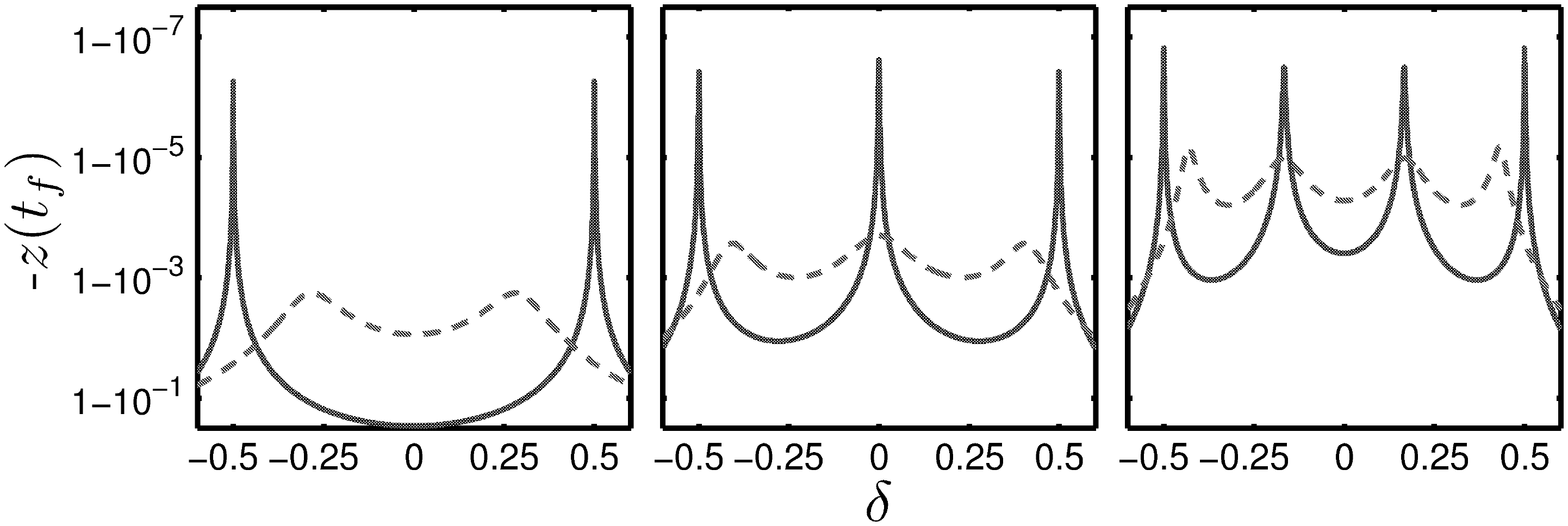}
\caption{\emph{From left to right:} Robustness profile of the time-optimal control pulse for the inversion of two spins with $\Delta_i\in\{-1/2,1/2\}$, three spins with $\Delta_i\in\{-1/2,0,1/2\}$ and four spins with $\Delta_i\in\{-1/2,-1/6,1/6,1/2\}$ (solid line), and of the control field derived with a GRAPE optimization (dashed line).\label{figA10}}
\end{figure}

We proceed as follows. We have determined the optimal time for controlling two spins with $\Delta_i\in\{-1/2,1/2\}$, three spins with $\Delta_i\in\{-1/2,0,1/2\}$ and four spins with $\Delta_i\in\{-1/2,-1/6,1/6,1/2\}$ (see Fig.~\ref{fig5}). We optimize a pulse of the same duration with GRAPE for 100 spins so that $\Delta\in[-1/2,1/2]$. A guess field is chosen for the optimization algorithm in order to recover the results of Fig. 5 of Ref.~\cite{kobzar1}. The result is presented in Fig.~\ref{figA9}.
We observe a strong similarity between the GRAPE solution and the global one with two, three and four spins. This result was not obvious since the GRAPE algorithm optimizes a large number of spins. Another interesting point is the robustness profile achieved with each pulse. We compute the fidelity $-z(t_f)$ obtained by integrating Eq.~\eqref{edEnsembleopt} for 1000 spins with an offset $\Delta\in[-0.6,0.6]$. The result is displayed in  Fig.~\ref{figA10}.
A remarkable point is the fact that the number of peaks of the robustness profiles is the same if a small or a large number of spins is considered. This suggests a way to limit the computational time of robust control sequences.
\section{Implementation of one-qubit quantum gates\label{SectGate}}
\begin{figure}[h!]
\centering
\includegraphics[angle=-90,scale=0.4]{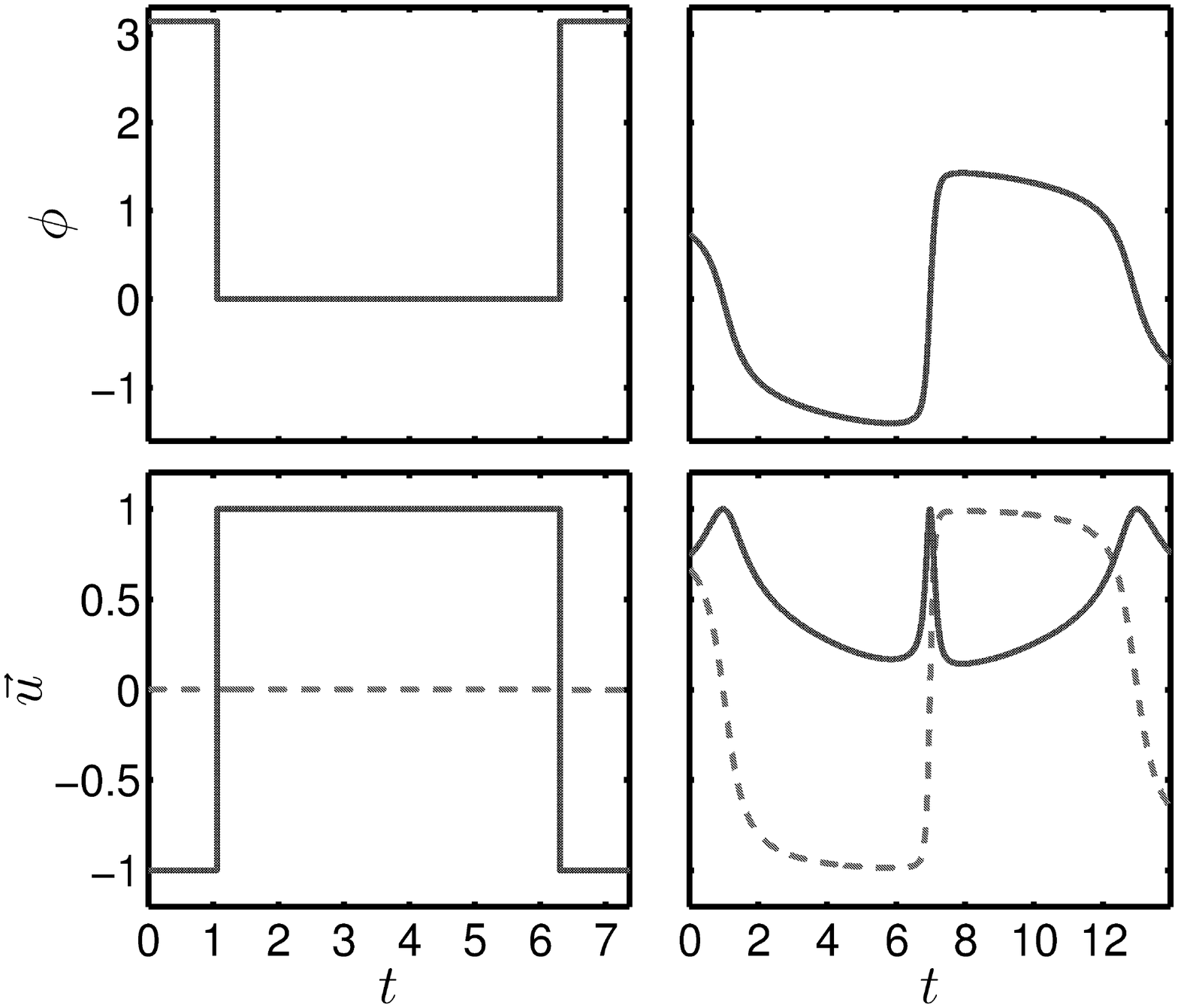}
\caption{\textit{Upper panels:} Phase $\Phi$ of the control fields which realizes a NOT gate robust to first order in $\delta$ (left), and to  second order (right). \textit{Lower panels:} Corresponding control fields $u_x=\cos\Phi$ and $u_y=\sin\Phi$. The second order solution is not exact, in the sense that the inhomogeneous part of the Bloch matrix is cancelled with a precision of the order of 0.1.\label{figA11}}
\end{figure}

The method presented in this work can be adapted to the implementation of quantum gates. A quantum gate can be written as a rotation matrix belonging to $SO(3)$. As an example, a NOT- gate is associated with a rotation of angle $\pi$ about the axis $\vec{e}_x$ (or $\vec{e}_y$) of the Bloch sphere. This matrix is given by:
\begin{equation}
G_{\text{NOT}}=\begin{pmatrix}
1 & 0 & 0\\0 & -1 & 0\\0 & 0 & -1
\end{pmatrix}.
\end{equation}
The Bloch equation can be derived for this rotation matrix, which is equivalent to consider the dynamics of three orthogonal Bloch vectors. This matrix will be called below, the Bloch matrix. In the case of offset inhomogeneities, the evolution of the Bloch matrix is governed by the following equation:
\begin{equation}
\dot{R}=\begin{pmatrix}
0& \delta & -u_y \\ -\delta & 0 & u_x \\ u_y & -u_x & 0
\end{pmatrix}R.
\label{eqBlochGate}
\end{equation}
We assume that the solution of the system~\eqref{eqBlochGate} can be expressed as a perturbative expansion in terms of $\delta$:
\begin{equation}
R(t)=R_0(t)+\delta R_1(t)+\cdots +\delta^NR_N(t)+O(\delta^{N+1}).
\end{equation}
The matrix $R_0$ is the homogeneous part of the solution. It can be shown that the dynamics of the system is of the form:
\begin{equation}
\frac{d}{dt}\begin{pmatrix}
R_0\\ R_1\\ R_2 \\ \vdots \\ R_N
\end{pmatrix}=\begin{pmatrix}
H_0 & 0 & 0 & \cdots & 0\\
\partial_{\delta} H & H_0 & 0 & & 0\\
0 & \partial_{\delta}H & H_0 & & 0\\
\vdots & & \ddots & \ddots & \vdots\\
0 & \cdots & 0 & \partial_{\delta}H & H_0
\end{pmatrix}\begin{pmatrix}
R_0\\ R_1\\ R_2 \\ \vdots \\ R_N
\end{pmatrix},
\label{edRi}
\end{equation}
with:
\begin{equation}
H_0=\begin{pmatrix}
0& 0 & -u_y \\ 0 & 0 & u_x \\ u_y & -u_x & 0
\end{pmatrix},\quad \partial_{\delta}H=\begin{pmatrix}
0& 1 & 0 \\ -1 & 0 & 0 \\ 0 & 0 & 0
\end{pmatrix}.
\end{equation}
At time $t=0$, the matrix $R_0(0)$ is the identity matrix and the matrices $R_{k\geq 1}$ are zero. A robust gate $G$ is a transfer of the form:
\begin{eqnarray}\label{GateRobust}
& & R_0(0)=\Id_{3\times 3}\mapsto R_0(t_f)=G,\\
& & R_{k\geq 1}(0)=0_{3\times 3}\mapsto R_{k\geq1}(t_f)=0_{3\times 3}.
\end{eqnarray}
Each matrix $R_k$ has nine components that we denote by $a^{(k)}_{ij}$, i.e.:
\begin{equation}
R_k=\begin{pmatrix}
a_{11}^{(k)} & a_{12}^{(k)} & a_{13}^{(k)}\\
a_{21}^{(k)} & a_{22}^{(k)} & a_{23}^{(k)}\\
a_{31}^{(k)} & a_{32}^{(k)} & a_{33}^{(k)}
\end{pmatrix}.
\end{equation}
We introduce the vector $\vec{q}_k$ (with nine-elements) defined as the concatenation of the three columns of the matrix $R_k$, i.e. $\vec{q}_k={^t(a_{11}^{(k)},a_{21}^{(k)},\cdots,a_{33}^{(k)})}$. The pseudo-Hamiltonian $H_P$ of the system is given by:
\begin{equation}
H_P=\sum_{k=1}^K\vec{p}_k\cdot\dot{\vec{q}}_k+p^0f^0,
\end{equation}
where $p^0f^0$ depends on the cost functional of the system. The components of the adjoint state $\vec{p}$ are given by:  $\vec{p}_k={}^t(b_{11}^{(k)},b_{21}^{(k)},\cdots,b_{33}^{(k)})$. We also define some three-dimensional angular momentum vectors $\vec{\ell}^{(i,j)}$ as:
\begin{equation}
\vec{\ell}^{(i,j)}=\begin{pmatrix}
\ell_{x}^{(i,j)}\\
\ell_{y}^{(i,j)}\\
\ell_{z}^{(i,j)}
\end{pmatrix}=\sum_{n=1}^3 \begin{pmatrix}
b_{1n}^{(i)}\\
b_{2n}^{(i)}\\
b_{3n}^{(i)}
\end{pmatrix}\times \begin{pmatrix}
a_{1n}^{(j)}\\
a_{2n}^{(j)}\\
a_{3n}^{(j)}
\end{pmatrix},
\end{equation}
and we introduce the vectors $\vec{\Omega}_k$ defined as:
\begin{equation}
\vec{\Omega}_k=\sum_{i=k}^N\vec{\ell}^{(i,i-k)},\quad k\in\{0,\cdots,N\}.
\label{defOmegakGate}
\end{equation}
Finally, we can show that the pseudo-Hamiltonian can be expressed in terms of the vectors $\vec{\Omega}_k$ as follows:
\begin{equation}
H_P=\vec{u}\cdot\vec{\Omega}_{0}+\Omega_{1z},
\end{equation}
with $\vec{u}=(u_x,u_y,0)$. Using the Hamilton's equations, we can deduce the dynamics of each $\vec{\Omega}_{k}$. We find exactly the same equation as Eq.~\eqref{edEopt}, that is:
\begin{equation}
\dot{\vec{\Omega}}_k=\vec{\Omega}_k\times\vec{u}+ \vec{\Omega}_{k+1}\times\vec{e}_z,\quad k\in\{0,\cdots,N\}.
\label{eqOmegaGate}
\end{equation}
We can then apply the PMP. For the time-minimum problem, the maximization of the pseudo-Hamiltonian leads to $u_x=\tfrac{\Omega_{0x}}{r}$ and $u_y=\tfrac{\Omega_{0y}}{r}$,
with $r=\sqrt{\Omega_{0x}^2+\Omega_{0y}^2}$. Substituting the expressions of $u_x$ and $u_y$ into Eq.~\eqref{eqOmegaGate}, it is then straightforward to show that $\Omega_{0z}$ is constant, but this constant can be different from zero. We set $\Omega_{0z}=I$. The time-optimal control field is solution of the following system of equations:
\begin{equation}
\begin{cases}
\dot{\vec{\Omega}}_0=\left(\vec{\Omega}_1-\tfrac{I}{r}\vec{\Omega}_0\right)\times\vec{e}_z\\
\dot{\vec{\Omega}}_k=\tfrac{1}{r}\vec{\Omega}_{k}\times\vec{\Omega}_0 +\left(\vec{\Omega}_{k+1}-\tfrac{I}{r}\vec{\Omega}_k\right)\times\vec{e}_z,\\\textrm{with}~k\in\{1,\cdots,N-1\},\\
\dot{\vec{\Omega}}_N=\tfrac{1}{r}\vec{\Omega}_{N}\times\vec{\Omega}_0-\tfrac{I}{r}\vec{\Omega}_N\times\vec{e}_z.
\end{cases}
\label{edfieldOptGate}
\end{equation}
Pontryagin's Hamiltonian is constant and is given by:
\begin{equation}
H=r+\Omega_{1z}.
\end{equation}
A difference from a state to state control problem is that, in general, $\Omega_{kz}\neq 0$. The dimension of the control landscape can only be reduced by one by noting that $|\vec{\Omega}_N|$ is a constant, which can be set to $1$ without loss of generality. Thus, a solution of the system~\eqref{edfieldOptGate} depends on $3N+2$ parameters which are $\Omega_{0x}(0)$, $\Omega_{0y}(0)$, $\Omega_{0z}=I$, $\Omega_{kx}(0)$, $\Omega_{ky}(0)$, $\Omega_{kz}(0)$, $\vartheta$ and $\varphi$,
with $\Omega_{Nx}(0)=\sin\vartheta\cos\varphi$, $\Omega_{Ny}(0)=\sin\vartheta\sin\varphi$ and $\Omega_{Nz}(0)=\cos\vartheta$. Figure~\ref{figA11} displays the control fields robust to first and second order in $\delta$. Further work will be necessary to improve this first solution.

\section{Conclusion}\label{conc}
Using the PMP, we have derived the global robust optimal control strategies (for the lowest orders) for the inversion transfer in the energy and time- minimum cases. The derived pulses have an explicit and relatively simple form which is easily implementable experimentally. The analytical expression of some control fields is also obtained. We stress that the global optimality of the solutions of this work is in sharp contrast with the fields designed by numerical methods, which correspond to local maxima.

These results can be viewed as a first step towards a complete answer of the robustness issue, which is a long-standing problem in quantum control. They also pave the way to other studies using the same approach, such as a transfer robust with respect to the two inhomogeneous parameters or the design of robust propagators, which will be interesting in quantum computing. A first step in this direction is made in Sec.~\ref{SectGate} with the derivation of a NOT gate robust to the second order in $\delta$. The main obstacle in these two generalizations will be the dimension of the control landscape, which makes the search for global optimal controls difficult. Another interesting issue would be to generalize this approach to the robustness against noise for which the experimental uncertainties are not constant in time.\\ \\
\noindent\textbf{ACKNOWLEDGMENT}\\
S.J. Glaser acknowledges support from the DFG (Gl 203/7-2). D. Sugny and S. J. Glaser acknowledge support from the ANR-DFG research program Explosys (ANR-14-CE35-0013-01). D. Sugny acknowledges support from the PICS program and from the ANR-DFG research program COQS (ANR-15-CE30-0023-01). The work of D. Sugny has been done with the support of the Technische Universit\"at M\"unchen – Institute for Advanced Study, funded by the German Excellence Initiative and the European Union Seventh Framework Programme under grant agreement 291763.
\appendix
\section{Application of the PMP in the energy minimum case}\label{appa}
We consider the control of the population inversion in the energy minimum case. We show how to determine the control field robust at order one, two and three. The general method consists in first computing a solution of the system~\eqref{edEopt}, and then using the control fields $u_x=\Omega_{0x}$ and $u_y=\Omega_{0y}$ in the system~\eqref{eq1} in order to realize a robust inversion. We add the constraint to use only one control field by setting $u_y=0$. Note that better results could be achieved at second order (but not at first order) when two fields are considered.
\subsection{Analytical derivation at first order}
\begin{figure}[h!]
\centering
\includegraphics[angle=-90,scale=0.2,trim=0 50 0 0,clip]{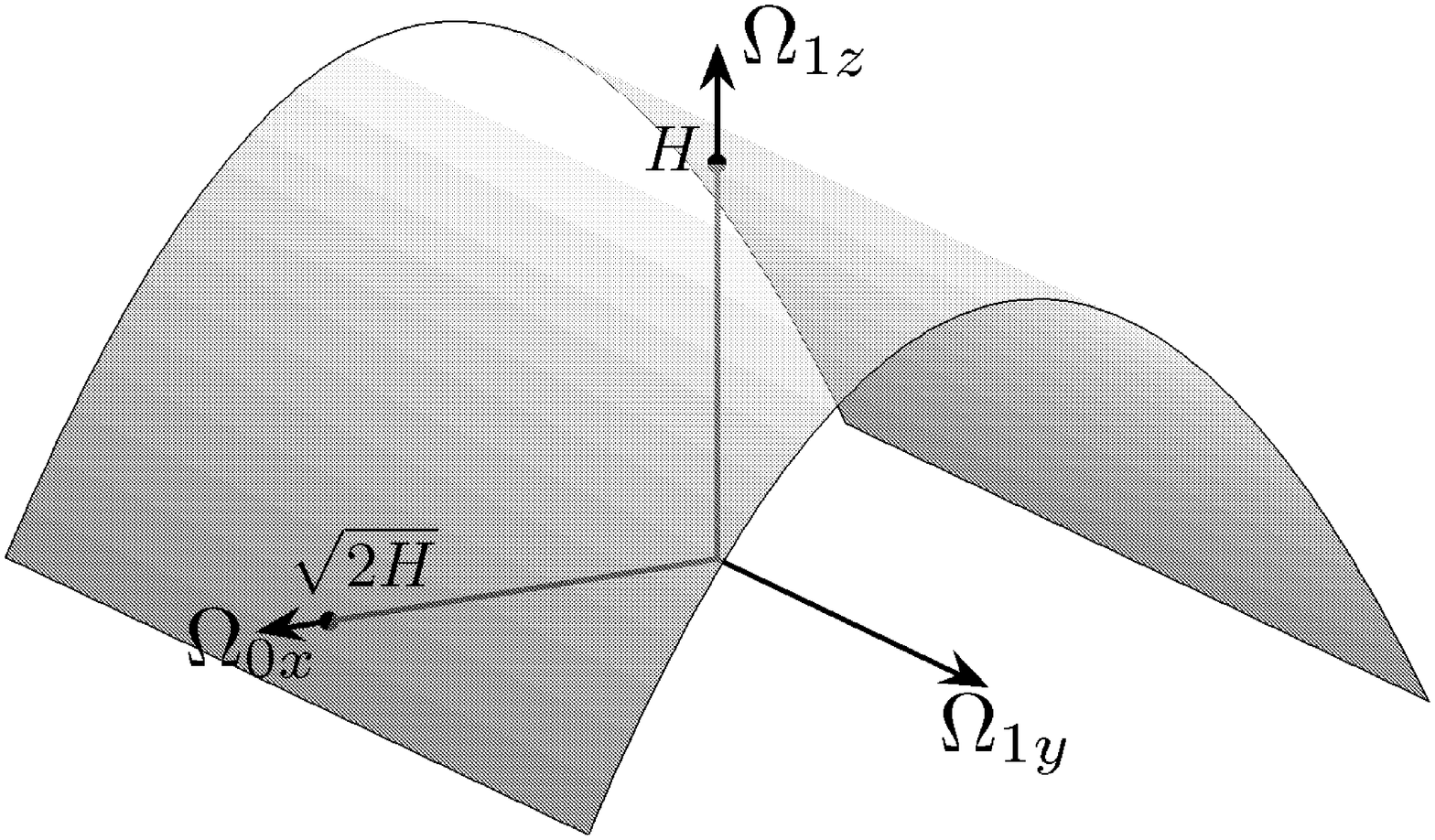}
\includegraphics[angle=-90,scale=0.2]{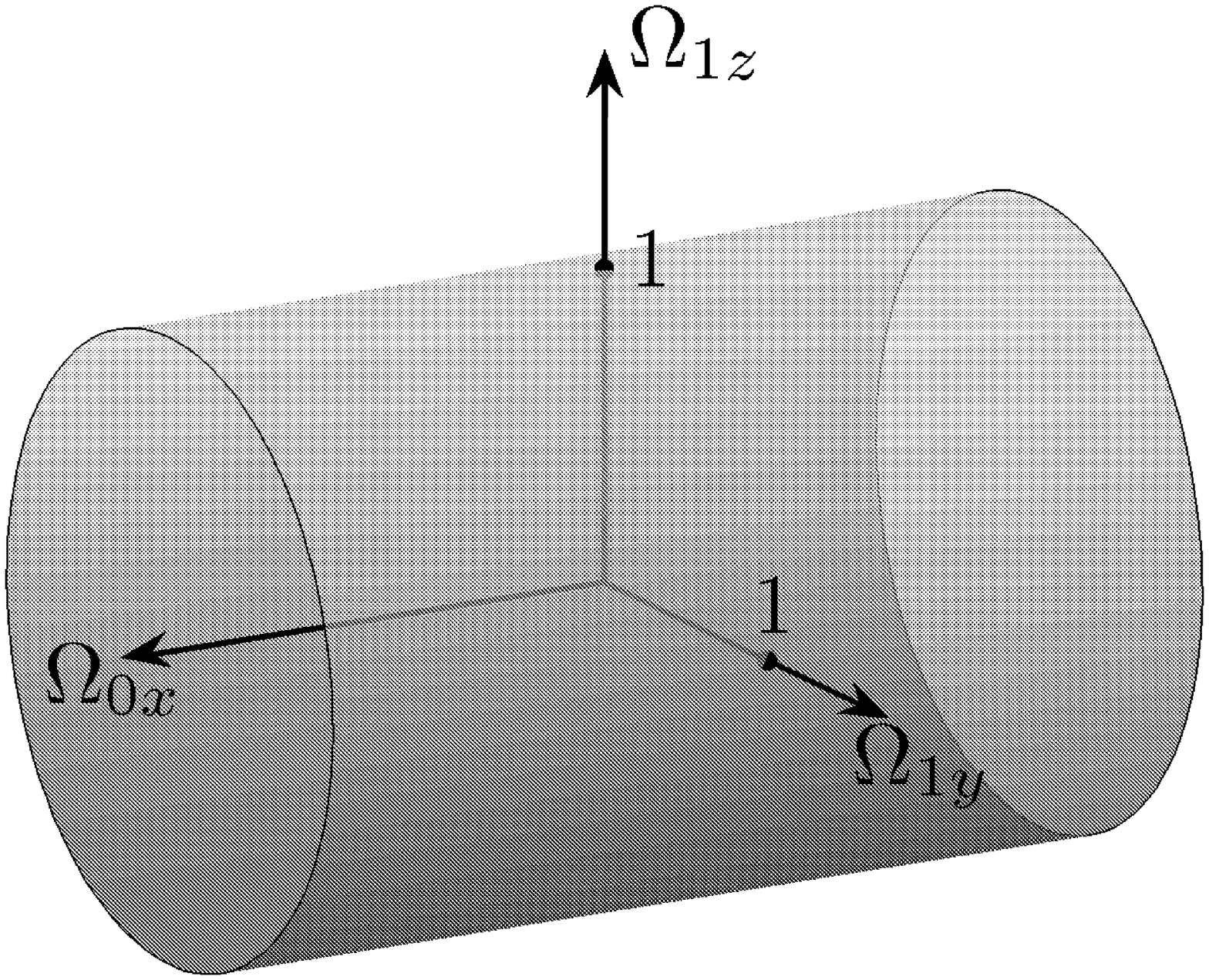}
\includegraphics[angle=-90,scale=0.2]{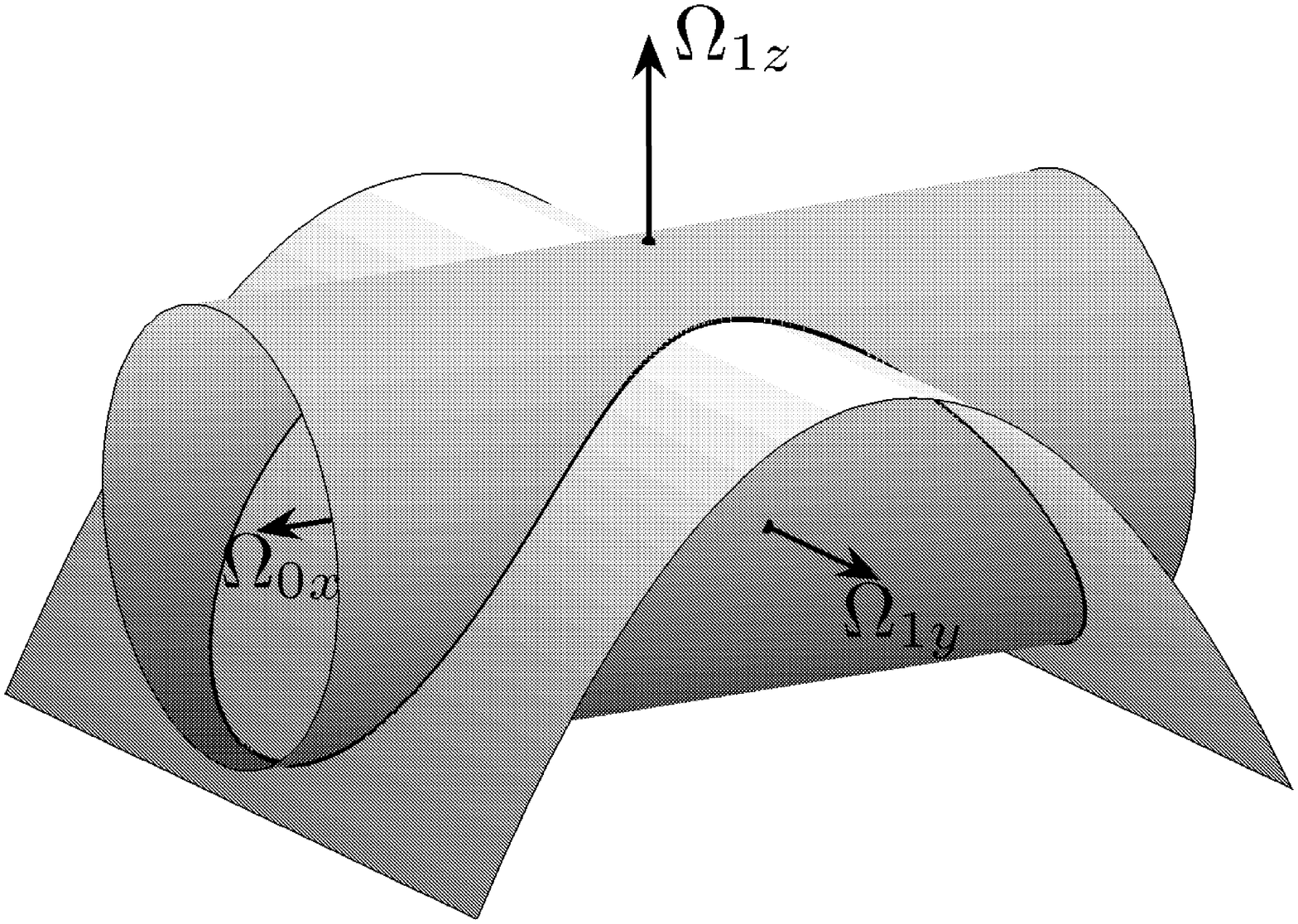}
\includegraphics[angle=-90,scale=0.2]{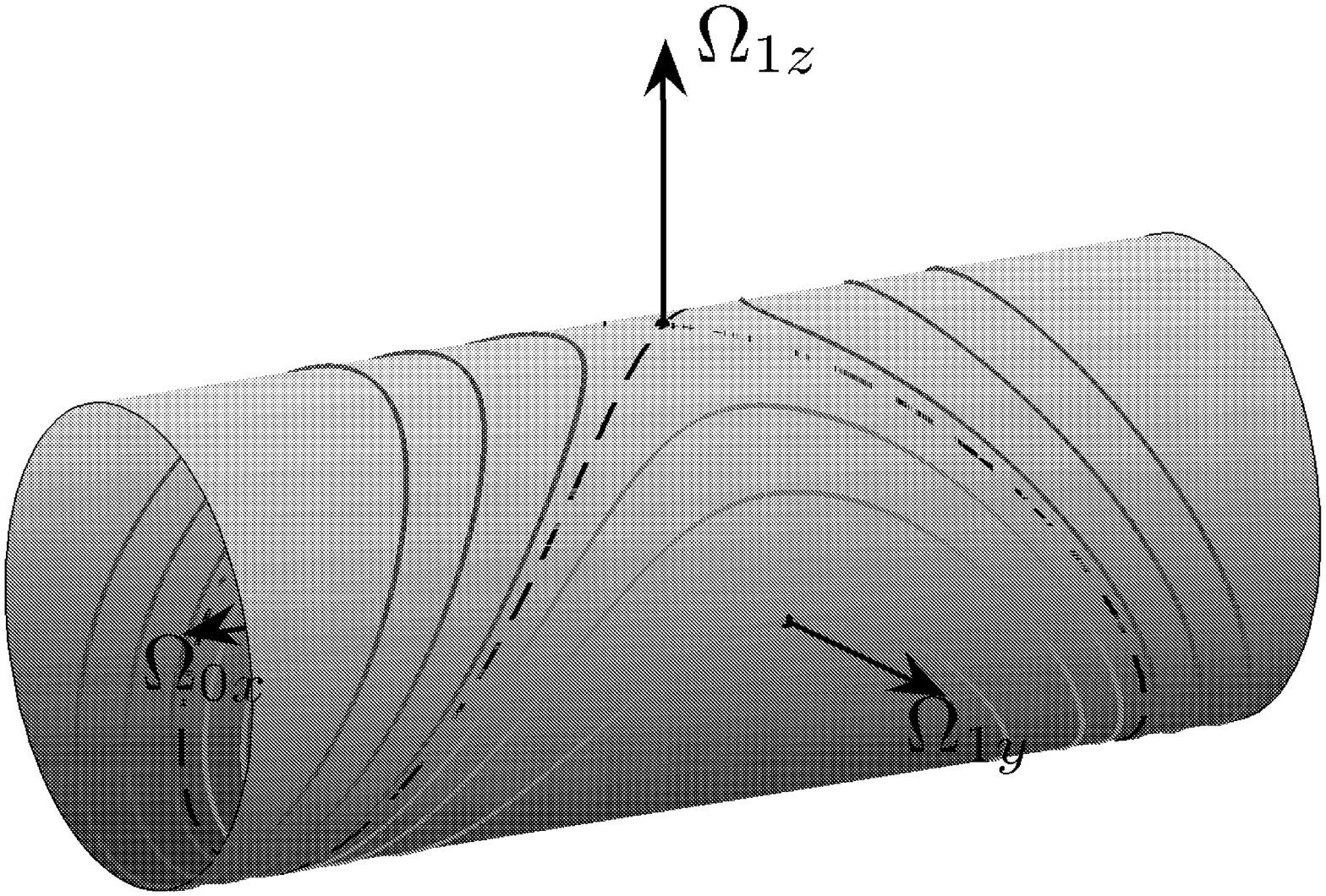}
\caption{\textit{Upper panels:} Parabolic plane associated with the conservation of $H$ (left), and the cylinder corresponding to the one of $\ell$ (right; see Eq.~\eqref{ConstDO1}).  \textit{Lower panels:} Intersection of the two surfaces for a given value of $H<1$ (left). The solution of the system~\eqref{systDeltaO1} belongs to this intersection (represented by a solid line). Phase portrait of the system~\eqref{systDeltaO1} plotted on the cylinder (right). The rotating and oscillating trajectories are respectively represented in dark gray and in light gray. The separatrix is the black dotted line.\label{figA1}}
\end{figure}

For $N=1$, the system of Eq.~\eqref{edEopt} becomes:
\begin{equation}
\dot{\vec{\Omega}}_0=\vec{\Omega}_1\times\vec{e}_z,\quad\dot{\vec{\Omega}}_1=\vec{\Omega}_1\times\vec{\Omega}_0.
\end{equation}
Setting $u_y=\Omega_{0y}=0$, we can show that the system simplifies to:
\begin{equation}
\dot{\Omega}_{0x}=\Omega_{1y},\quad\dot{\Omega}_{1y}=\Omega_{0x}\Omega_{1z},\quad \dot{\Omega}_{1z}=-\Omega_{0x}\Omega_{1y}.
\label{systDeltaO1}
\end{equation}
It has two constants of motion:
\begin{equation}
H=\tfrac{1}{2}\Omega_{0x}^2+\Omega_{1z},\quad \ell=\Omega_{1y}^2+\Omega_{1z}^2.
\label{ConstDO1}
\end{equation}
We can set $\ell=1$ without loss of generality. These two constants are associated with two surfaces in the $(\Omega_{0x},\Omega_{1y},\Omega_{1z})$- space. The conservation of $H$ corresponds to a parabolic plane, and the one of $\ell$ to a cylinder of radius $\ell=1$ along the $\Omega_{0x}$- direction. These two surfaces are represented in Fig.~\ref{figA1}. The solution of the system~\eqref{systDeltaO1} belongs to the intersection of these two surfaces. The plot of this intersection for different values of $H$ leads to the phase portrait of the system~\eqref{systDeltaO1} depicted on the cylinder. We get the same phase portrait as the one of a planar pendulum, with three families of solution \cite{goldstein}. The \emph{rotating solutions} occur for $H>1$, the \emph{oscillating solutions} for $-1\leq H\leq 1$, and the separatrix for $H=1$. The trajectories for which $-1\leq H < 0$ are not solutions of the problem, since we have in this case $\Omega_{1z}<0$, which is not possible since $\Omega_{1z}(0)=0$. 
The solution of the system~\eqref{systDeltaO1} can be expressed in terms of the Jacobi amplitude function $\am(u,m)$~\cite{Abrambook}, which is an angle defined as the inverse of the Incomplete Elliptic Integral of the first kind $\Fell(u,m)=\int_{0}^{u}\frac{dt}{\sqrt{1-m\sin^2t}}$. Note the relation $d\am(u,m)/du=\sqrt{1-m\sin^2(\am(u,m))}$ which is useful for integrating the following equations. We only consider the oscillating trajectories which contain the global optimum of the problem. The explicit solution of the system~\eqref{systDeltaO1} can be expressed in the oscillating case as $\Omega_{0x}=2\sqrt{m}\cos\nu$, $\Omega_{1y}=-2\sqrt{m}\sin\nu\sqrt{1-m\sin^2\nu}$ and $\Omega_{1z}=2m\sin^2\nu-1$, where
$\nu(t)=\am(t+\rho,m)$, $\rho=\pm\Fell\left(\arcsin\left(\tfrac{1}{\sqrt{2m}}\right),m\right)$ and $m=\tfrac{1+H}{2}$. The initial phase $\rho$ is computed so as $\Omega_{10z}=0$. Note that the initial conditions are so that $\Omega_{0x}(0)=\sqrt{2H}$ and $\Omega_{1y}(0)=\pm 1$, the sign being the same as the sign of $-\rho$.

\begin{figure}[h!]
\includegraphics[angle=-90,scale=0.4]{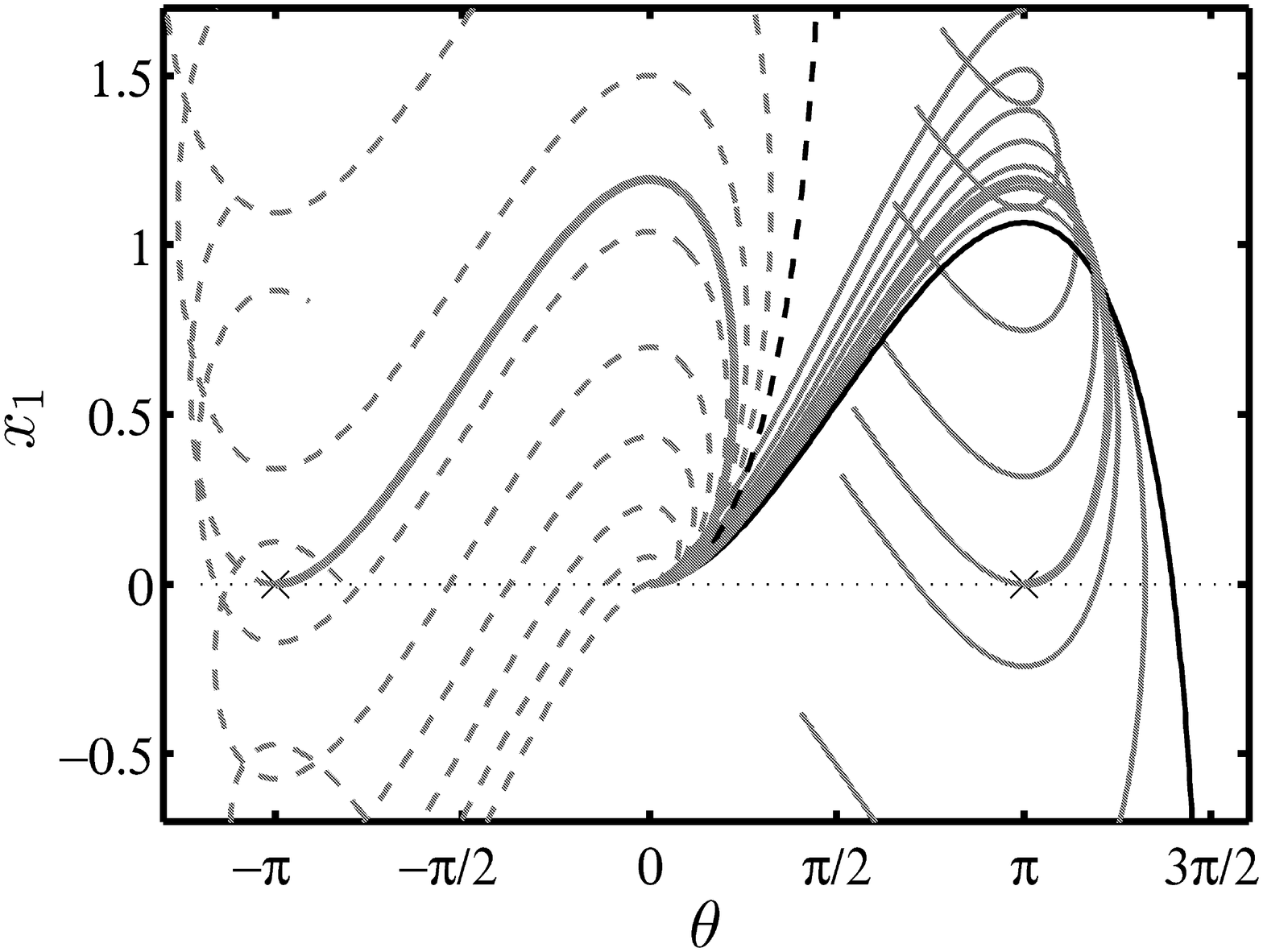}
\centering
\caption{Plot of the trajectories $x_1$ as a function of the angle $\theta$. The dashed lines correspond to $\rho=\Fell\left(\arcsin\left(\tfrac{1}{\sqrt{2m}}\right),m\right)$ ($\Omega_{1y}(0)=-1$), and the solid lines to $\rho=-\Fell\left(\arcsin\left(\tfrac{1}{\sqrt{2m}}\right),m\right)$ ($\Omega_{1y}(0)=1$). The crosses represent the robust target sate $(\theta(t_f),x_1(t_f))= (\pm\pi,0)$. The black lines are the solutions associated with the separatrices of Fig.~\ref{figA1}, that is for the limit $H\to 1$. The thick brown lines are two equivalent optimal solutions for the population inversion. \label{figA2}}
\end{figure}

Each trajectory of the phase portrait is associated with a control $\Omega_{0x}(t)$ which is a candidate to realize a robust population inversion. Since a control field depends only on $H$ (and eventually on the sign of $\rho$), the goal is to find the parameter $H$ which inverts the population with a minimum of energy. For that purpose, we integrate the coordinates $\vec{q}_i(t)$. At first order in $\delta$, the differential system governing the dynamics of $\vec{q}_0=(x_0,y_0,z_0)$ and $\vec{q}_1=(x_1,y_1,z_1)$ is given by:
\begin{equation}
\dot{\vec{q}}_0=\vec{q}_0\times\vec{\Omega}_0,\quad \dot{\vec{q}}_1=\vec{q}_0\times\vec{e}_z+ \vec{q}_1\times\vec{\Omega}_0.
\end{equation}
Since we have $\Omega_{0y}(t)=\Omega_{0z}(t)=0$ and $\vec{q}_1(0)=(0,0,0)$, the system simplifies into:
\begin{equation}
\dot{y}_0=\Omega_{0x}z_0,\quad \dot{z}_0=-\Omega_{0x}y_0,\quad \dot{x}_1=y_0.
\end{equation}
All the other coordinates are equal to zero. Introducing the angle $\theta(t)$ so that $\sin\theta=y_0$ and $\cos\theta=z_0$, the problem consists in computing the parameter $H$ which realizes the transfer:
\begin{equation}
(\theta(0),x_1(0))=(0,0)\rightarrow(\theta(t_f),x_1(t_f))= (\pm\pi,0).
\end{equation}
The differential system is given in these coordinates by:
\begin{equation}
\dot{\theta}=\Omega_{0x},\quad\dot{x}_1=\sin\theta.
\end{equation}
The solution of this system can be written as:
\begin{equation}
\theta(t)=2\arcsin(\sqrt{m}\sin\nu(t))\pm\tfrac{\pi}{2},\quad x_1(t)=\int_0^{t}\sin\theta(t') dt',
\end{equation}
where the sign is the same as the one of $\rho$. The solution $x_1(t)$ could be expressed as the sum of a linear term and an Elliptic Integral of the second kind \cite{Abrambook}. Figure~\ref{figA2} shows the trajectories $x_1=f(\theta)$ starting from $(0,0)$ for different values of $H$.
Note that Fig.~\ref{figA2} represents some trajectories starting from the north pole of the Bloch sphere driven by the control fields solution of the PMP. Optimal robust transfers can be found from Fig.~\ref{figA2}. As an example, it can be seen that the point $(\theta=-\pi/2,x_1=0)$ belongs to some of the trajectories, which corresponds to an optimal robust excitation transfer.

For the inversion, the optimal time $t^*$ is computed so that $\theta(t^*)=\pm\pi$. We find:
\begin{equation}
t^*=2\K(m)=2\int_{0}^{\tfrac{\pi}{2}}\frac{dt}{\sqrt{1-m\sin^2t}},
\end{equation}
where $\K$ is a complete elliptic integral of the first kind \cite{Abrambook}. A robust control is achieved by finding the parameter $H$ for which $\int_0^{t^*}\sin\theta dt=0$, in order to cancel the contribution to the first order term in $\delta$. We get:
\begin{equation}
H=0.6522,
\end{equation}
for the two possible signs of $\rho$. The corresponding trajectories in the $(\theta,x_1)$- plane are represented in Fig.~\ref{figA2}, for $\rho>0$ and $\rho<0$. The two optimal solutions are equivalent and give the global optimum. The corresponding pulse $u_x=\Omega_{0x}$ is plotted in Fig. 1. Its area is given by:
\begin{equation}
A=\int_0^{t^*}|\Omega_{0x}(t)|dt=1.45\pi,
\end{equation}
Note that other local optima exist and can be used to realize the population inversion.
\subsection{Derivation of the control field at second order\label{sect-O2E}}
\begin{figure}[h!]
\centering
\includegraphics[angle=-90,scale=0.25]{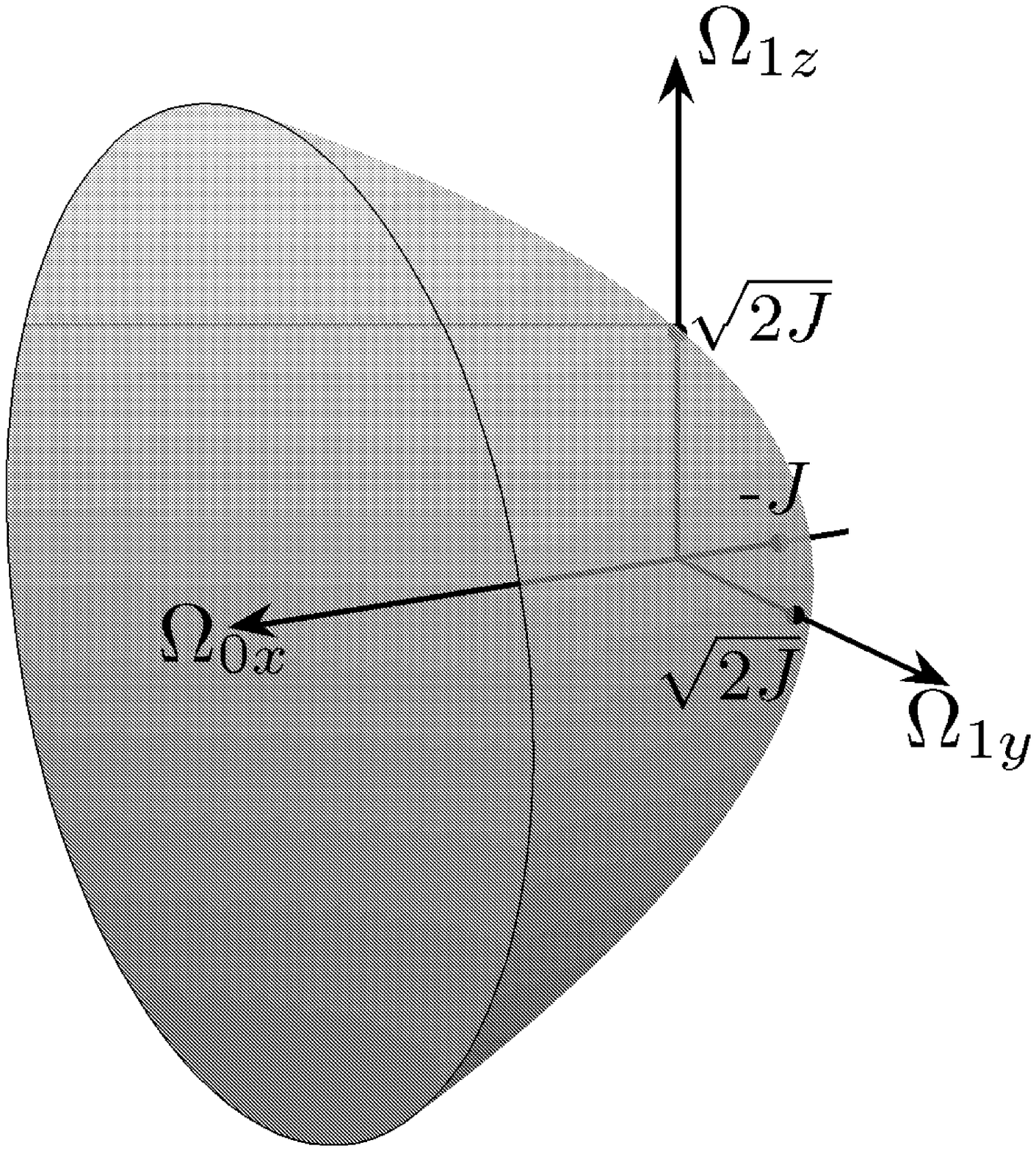}
\includegraphics[angle=-90,scale=0.25]{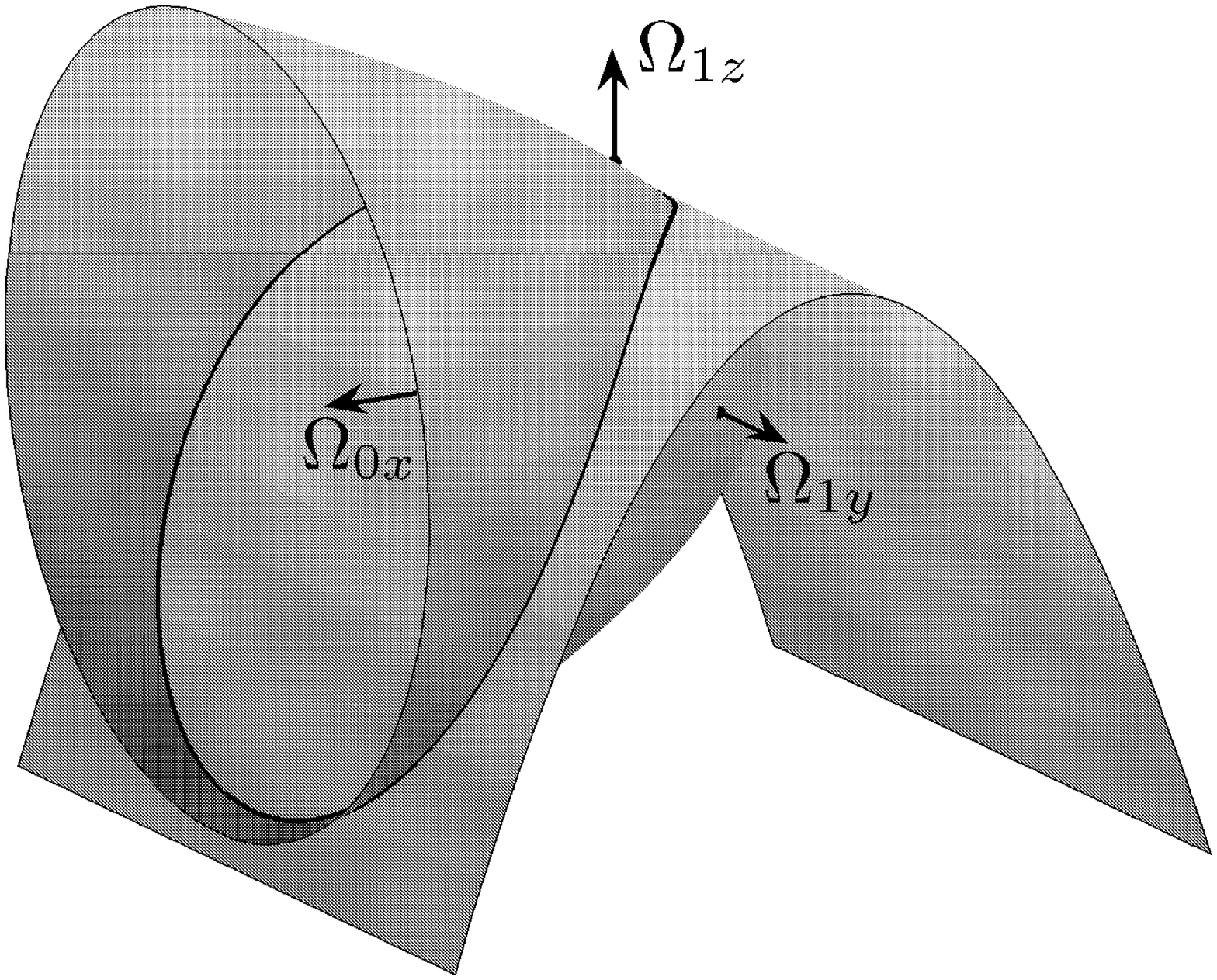}
\includegraphics[angle=-90,scale=0.225]{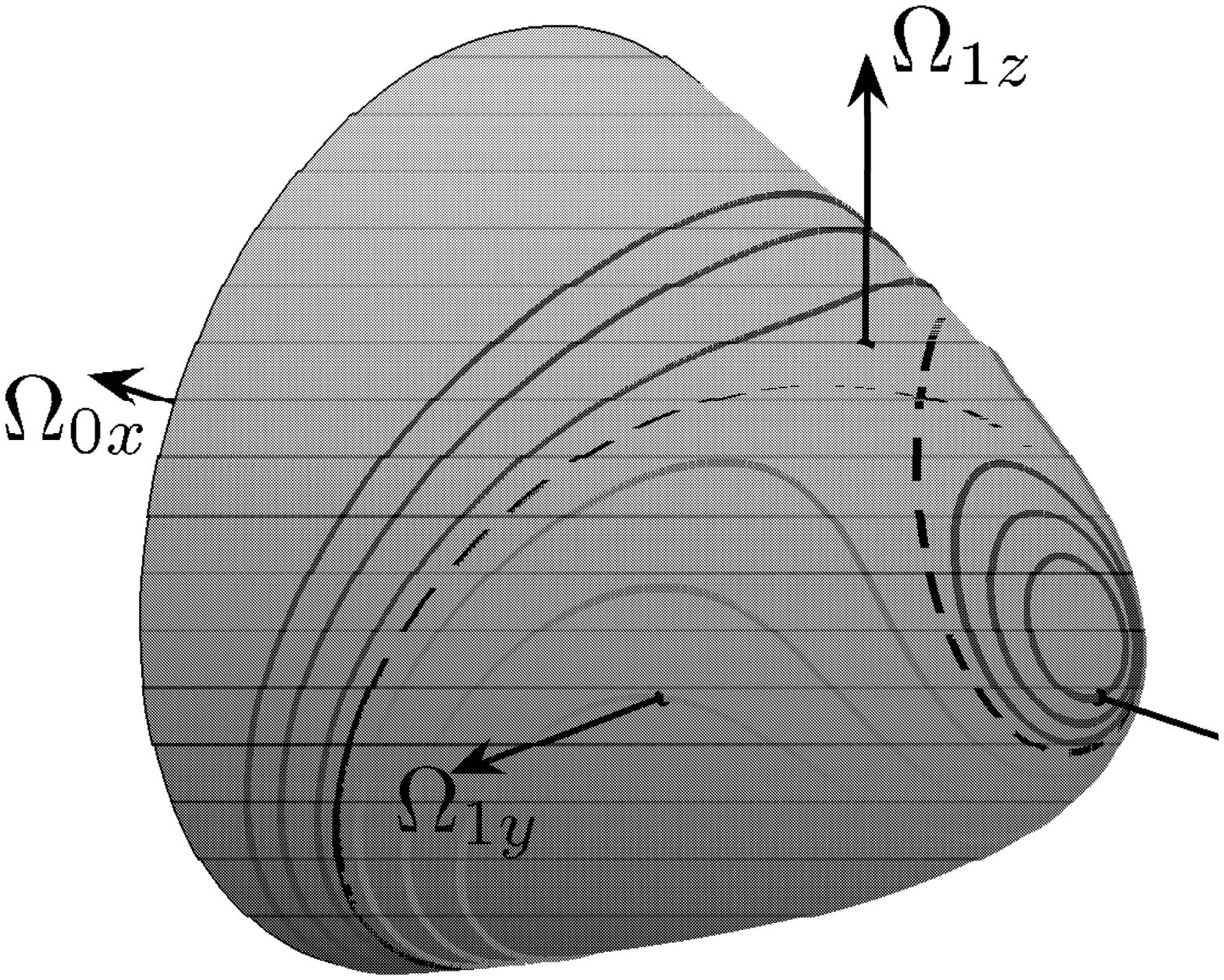}
\includegraphics[angle=-90,scale=0.225]{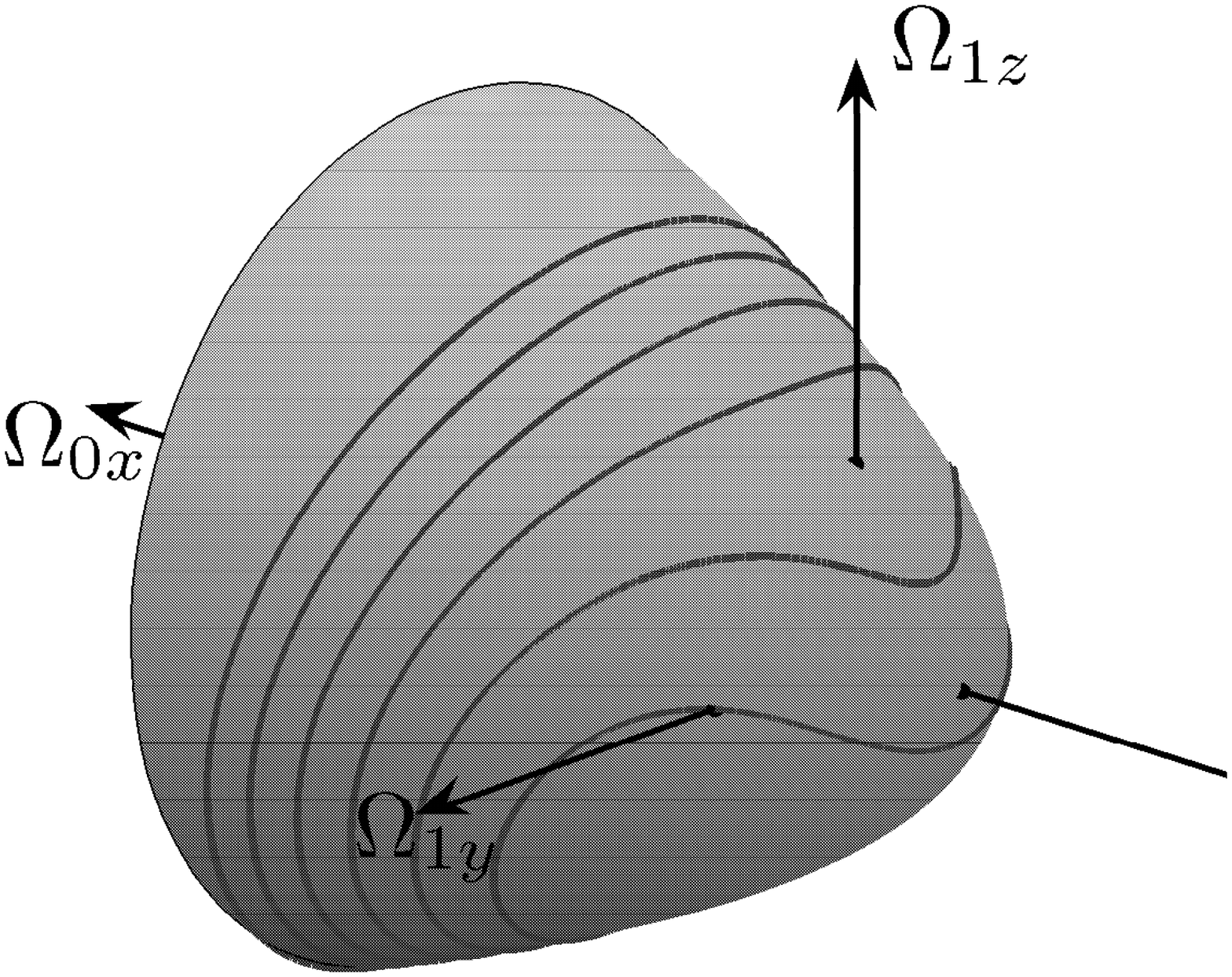}
\caption{\textit{Upper panels:} Paraboloid of $J$ constant. Intersection with the surface of $H$ constant (see Fig.~\ref{figA1}). The solution of the system~\eqref{ed1fEO2} belongs to this intersection, represented with a black curve. \textit{Lower panels:} Phase portrait of Eq.~\eqref{ed1fEO2} for two different values of $J$.\label{figA3}}
\end{figure}

We compute in this paragraph the solution at second order with one control field $u_x$. Better solutions can be achieved if a second field is considered. 
The second order solution satisfies the following differential system:
\begin{equation}
\dot{\vec{\Omega}}_0=\vec{\Omega}_1\times\vec{e}_z,~\dot{\vec{\Omega}}_1=\vec{\Omega}_1\times\vec{\Omega}_0+ \vec{\Omega}_2\times\vec{e}_z,~\dot{\vec{\Omega}}_2=\vec{\Omega}_2\times\vec{\Omega}_0.
\end{equation}
This system has six constants of the motion given by:
\begin{equation}
\begin{aligned}
&H=\tfrac{1}{2}\left|\vec{\Omega}_0\right|^2 +\Omega_{1z},\\
&I=\vec{\Omega}_1\cdot\vec{\Omega}_2,\\
&J=\tfrac{1}{2}\left|\vec{\Omega}_1\right|^2 +\vec{\Omega}_0\cdot\vec{\Omega}_2,\\
&K=\vec{\Omega}_0\cdot\vec{\Omega}_1+\Omega_{2z},\\
&0=\Omega_{0z},\\
&1=\left|\vec{\Omega}_2\right|^2.
\end{aligned}
\end{equation}
In the case $u_y=\Omega_{0y}=0$, we obtain that $K=I=0$ and that $\Omega_{1x}=\Omega_{2y}=\Omega_{3y}=0$, which also means that $\Omega_{2x}=-1$ (note that the sign could be set to $+1$, but the result is equivalent). 
The system then becomes:
\begin{equation}
\dot{\Omega}_{0x}=\Omega_{1y},\quad \dot{\Omega}_{1y}=\Omega_{0x}\Omega_{1z}+1,\quad \dot{\Omega}_{1z}=-\Omega_{1y}\Omega_{0x}.
\label{ed1fEO2}
\end{equation}
It has two constants of motion:
\begin{equation}
H=\tfrac{1}{2}\Omega_{0x}^2+\Omega_{1z},\quad J=\tfrac{1}{2}\Omega_{1y}^2+\tfrac{1}{2}\Omega_{1z}^2-\Omega_{0x}.
\end{equation}
The conservation of $H$ is associated with the same parabolic plane as the one represented in Fig.~\ref{figA1}. The conservation of $J$ is described by a symmetric paraboloid of axis $\Omega_{0x}$ as shown in Fig.~\ref{figA3}. The solution belongs to the intersection of these two surfaces, and the phase portrait is obtained by drawing it for every pair $(H,J)$. We plot the phase portrait for two different values of $J$ in Fig.~\ref{figA3}.
The system~\eqref{ed1fEO2} can be integrated by using the fact that:
\begin{equation}
\left(\dot{\Omega}_{0x}\right)^2=-\frac{1}{4}\Omega_{0x}^4+H\Omega_{0x}^2+2\Omega_{0x}+2J-H.
\end{equation}
The solution $\Omega_{0x}$ can be expressed in terms of Jacobi's elliptic functions~\cite{Joyeux96}. However, we use here a numerical analysis to find the optimal pulse.


A solution of the system~\eqref{ed1fEO2} depends on two parameters which are the initial conditions $\Omega_{0x}(0)$ and $\Omega_{1y}(0)$ ($\Omega_{1z}(0)$ is equal to zero by construction). These parameters can be related to $H$ and $J$ through the relations $\Omega_{0x}=\sqrt{2H}$ and $\Omega_{0y}(0)=\sqrt{2J+2\sqrt{2H}}$. In other words, the control landscape is two-dimensional and is parameterized by $H$ and $J$.

\begin{figure}[h!]
\centering
\includegraphics[angle=-90,scale=0.3,trim=10 0 0 0,clip]{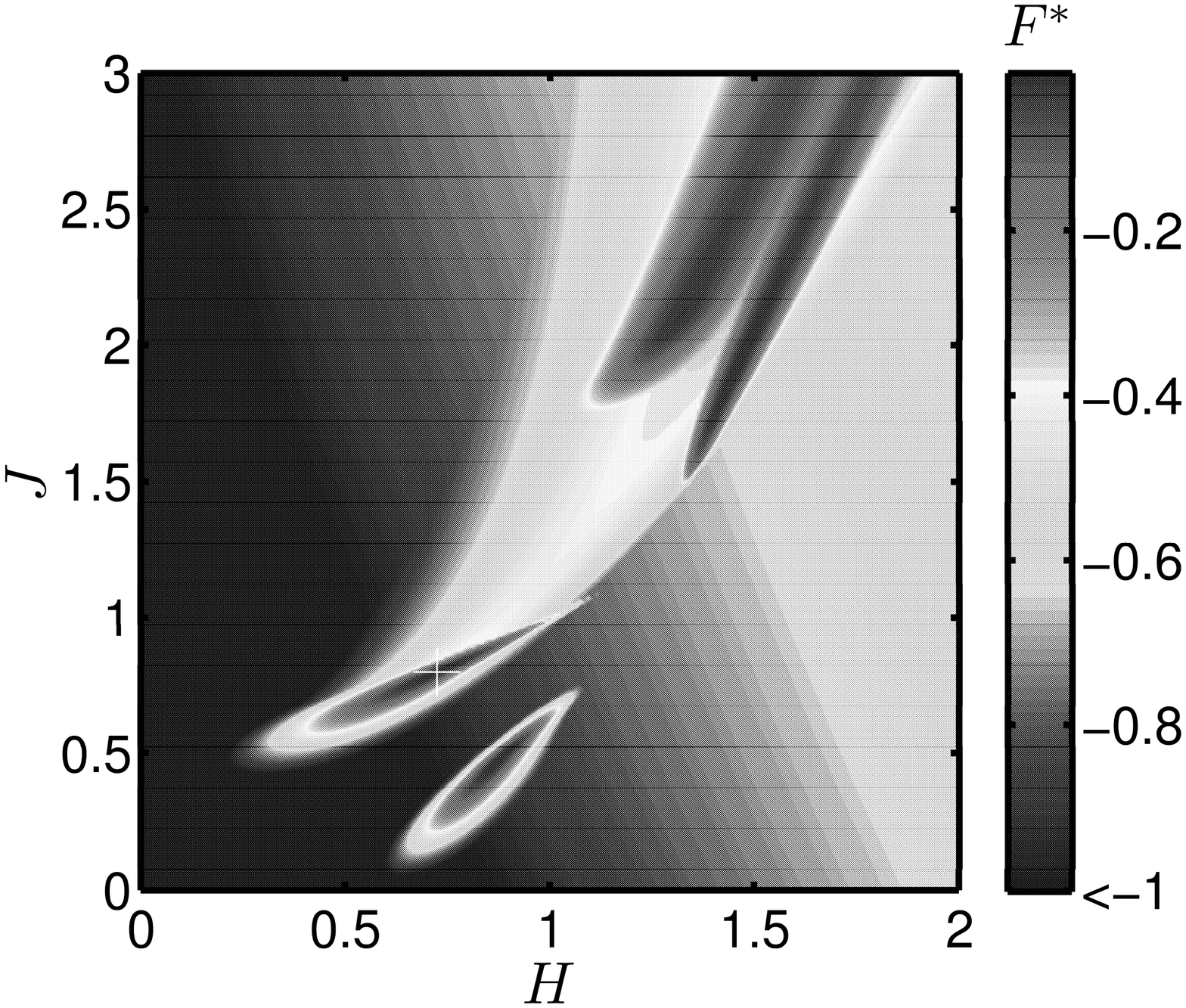}
\includegraphics[angle=-90,scale=0.3,trim=10 0 15 0,clip]{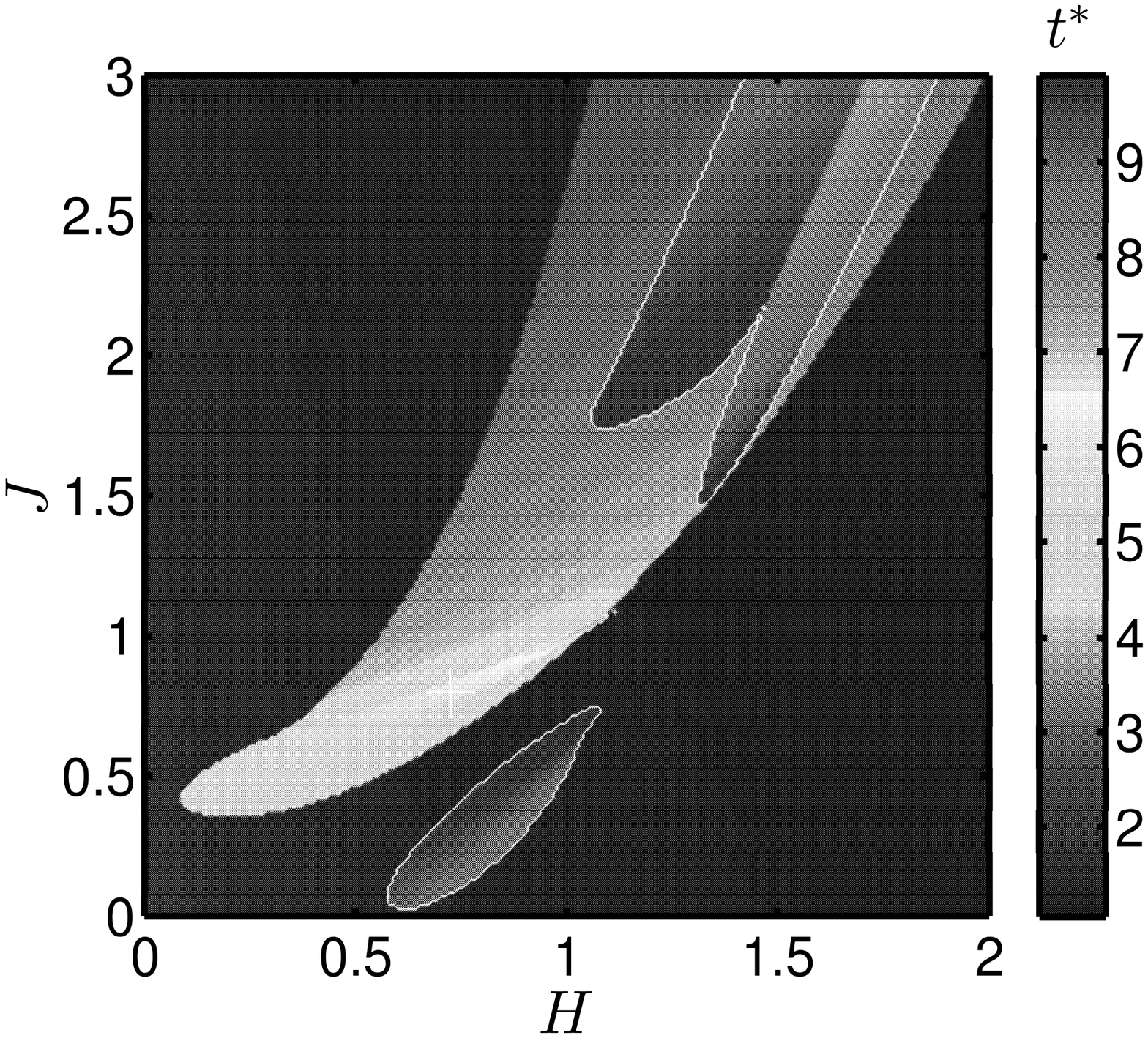}
\includegraphics[angle=-90,scale=0.3,trim=10 0 15 0,clip]{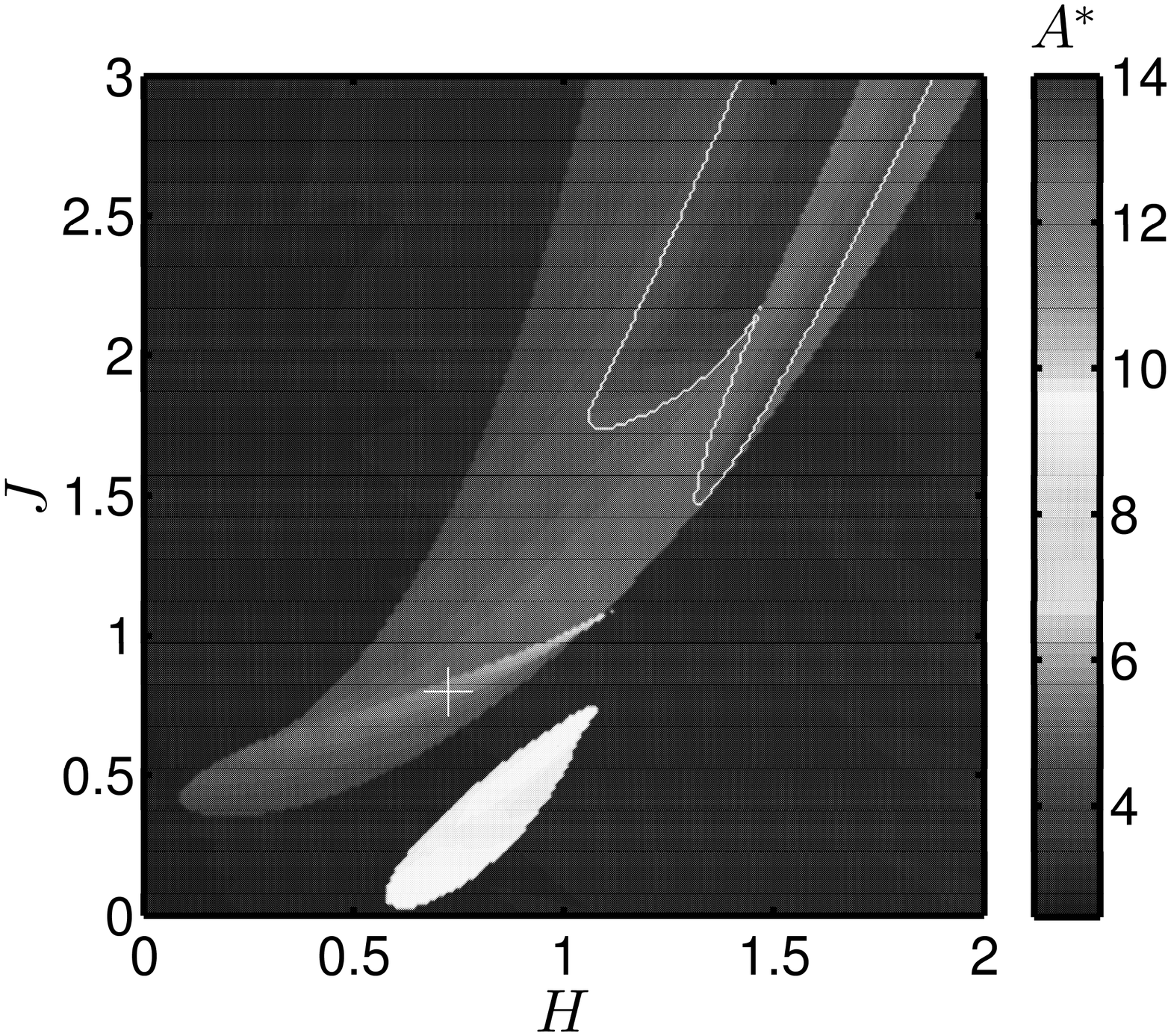}
\caption{Maximum value $F^*$ of the fidelity $F(t)$ of Eq.~\eqref{fidEO2} (top), time $t^*$ for which $F(t^*)=F^*$ (middle), and area of the corresponding control field (bottom) in the plane $(H,J)$. The white cross indicates the position of the global optimal solution.\label{figA4}}
\end{figure}

We introduce the following time-dependent fidelity:
\begin{equation}\label{fidEO2}
F(t)=-||\vec{q}_{0T}-\vec{q}_0(t)||^2-||\vec{q}_{1T}-\vec{q}_1(t)||^2-||\vec{q}_{2T}-\vec{q}_2(t)||^2,
\end{equation}
where $\vec{q}_{0T}=(0,0,-1)$ and $\vec{q}_{kT,k>0}=(0,0,0)$ are the target states for a robust population inversion. The general method can be described as follows. For a set of parameters $(H,J)$, we integrate numerically the system~\eqref{ed1fEO2} until an arbitrary time $t_f$. We then  use the solution $\Omega_{0x}(t)$ as a control field in the system~\eqref{eq1}. We integrate this system until the time $t=t_f$ and we compute the fidelity $F(t)$ of Eq.~\eqref{fidEO2}. This leads to the time $t^*$ for which the fidelity $F(t)$ is maximum, and we denote by $F(t^*)=F^*$ the corresponding fidelity. We also obtain the area of the control field $A^*=\int_{0}^{t^*}|\Omega_{0x}(t)|dt$.
Since the control landscape is a two-dimensional space, the different quantities $F^*$, $t^*$ and $A^*$ can be determined for every couple $(H,J)$. It is then straightforward to find the global optimal solution of the control problem. This approach is shown in Fig.~\ref{figA4}.
It can be seen in Fig.~\ref{figA4} that $F^*=0$ is satisfied for many pairs $(H,J)$. One solution is associated with the smallest energy (and with a very small area $A^*$). This solution is the global optimum of the control problem, and is given by:
\begin{equation}
H=0.7256,\quad J=0.7985,\quad t^*=1.95\pi.
\end{equation}
From a numerical point of view, the target is reached with a precision of $F^*=-3.4\times 10^{-6}$, and the area of the pulse is $A^*=1.81\pi$. The control field $u_x=\Omega_{0x}$ and the homogeneous solution $\vec{q}_0$ are represented in Fig.~1.
\subsection{Derivation of the control field at third order\label{sect-O3E}}
With only one control field $u_x$, we can show that the system governing the optimal control field is such that $\Omega_{0y}=\Omega_{1x}=\Omega_{2y}=\Omega_{2z}=\Omega_{3x}=0$. The other coordinates satisfy:
\begin{equation}
\begin{cases}
\dot{\Omega}_{0x}=\Omega_{1y},\\
\dot{\Omega}_{1y}=\Omega_{0x}\Omega_{1z}-\Omega_{2x},\\
\dot{\Omega}_{1z}=-\Omega_{0x}\Omega_{1y},\\
\dot{\Omega}_{2x}=\Omega_{3y},\\
\dot{\Omega}_{3y}=\Omega_{0x}\Omega_{3z},\\
\dot{\Omega}_{3z}=-\Omega_{0z}\Omega_{3y}.
\end{cases}
\label{edO3E}
\end{equation}
The system has the four following first integrals:
\begin{equation}
\begin{cases}
2H=\Omega_{0x}^2+2\Omega_{1z},\\
2I=\Omega_{2x}^2+2\Omega_{1y}\Omega_{3y}+2\Omega_{1z}\Omega_{3z},\\
2J=\Omega_{1y}^2+\Omega_{1z}^2+2\Omega_{0x}\Omega_{2x}+2\Omega_{3z},\\
1\equiv \Omega_{3y}^2+\Omega_{3z}^2.
\end{cases}
\end{equation}
Since the Bloch vector starts on the north pole of the sphere at time $t=0$, we have $\Omega_{kz}(0)=0$ for all $k$. We thus have $\Omega_{3y}(0)=-1$ (we could also choose $+1$), and a solution of the system~\eqref{edO3E} depends on the three following parameters:
\begin{equation}
\Omega_{0x}(0),\quad \Omega_{1y}(0),\quad \Omega_{2x}(0),
\end{equation}
which involves that the control landscape is $3$-dimensional.
An analytical solution of this system is difficult to compute. The corresponding parameters for a robust inversion are determined with a numerical gradient algorithm. We find:
$\Omega_{0x}(0)=1.2384$, $\Omega_{1y}(0)=2.9848$, $\Omega_{2x}(0)=-2.8019$ with and $t^*=2.43\pi$ and $F^*=-9\times 10^{-8}$. The area of the control field is:
\begin{equation}
A^*=\int_0^{t^*}|u_x(t)|dt=2.11\pi.
\end{equation}
The control fields and the homogenous solution $\vec{q}_0(t)$ are depicted in Fig.~1.
\section{Application of the PMP to the time-minimum case}\label{appb}
In this section, we apply the PMP to the time-minimum case.

For the optimal control problem at first order, the differential system can be expressed as:
\begin{equation}
\begin{cases}
\dot{\Omega}_{0x}=\Omega_{1y},\\
\dot{\Omega}_{0y}=-\Omega_{1x},\\
\dot{\Omega}_{1x}=-\tfrac{1}{r}\Omega_{0y}\Omega_{1z},\\
\dot{\Omega}_{1y}=\tfrac{1}{r}\Omega_{0x}\Omega_{1z},\\
\dot{\Omega}_{1z}=\left.\left.\tfrac{1}{r}\right(\Omega_{0y}\Omega_{1x}-\Omega_{0x}\Omega_{1y}\right).\\
\end{cases}
\label{edOmegaDeltaO1}
\end{equation}
In addition to the Pontryagin's Hamiltonian, this system has two first integrals of the form $|\vec{\Omega}_1|\equiv 1$ and $I=\vec{\Omega}_0\cdot\vec{\Omega}_1$. The control landscape of the system~\eqref{edOmegaDeltaO1} is parameterized by $\Omega_{0x}(0)$ and $\vartheta$ so that $\Omega_{1x}(0)=\cos\vartheta$ and $\Omega_{1y}(0)=\sin\vartheta$. A numerical analysis shows that the global optimum occurs for $\vartheta=\pi/2$. Since the quantity $I=\vec{\Omega}_0\cdot\vec{\Omega}_1$ is constant, and is equal to zero in this optimal situation, we deduce that $\vec{\Omega}_1$ is perpendicular to $\vec{\Omega}_0$ for any time $t$. This constraint can be fulfilled if and only if $\Omega_{0y}(t)=0$, that is $u_y(t)=0$. The control is therefore of the form $u_x=\sign(\Omega_{0x})$.

\begin{figure}[h!]
\centering
\includegraphics[angle=-90,scale=0.3]{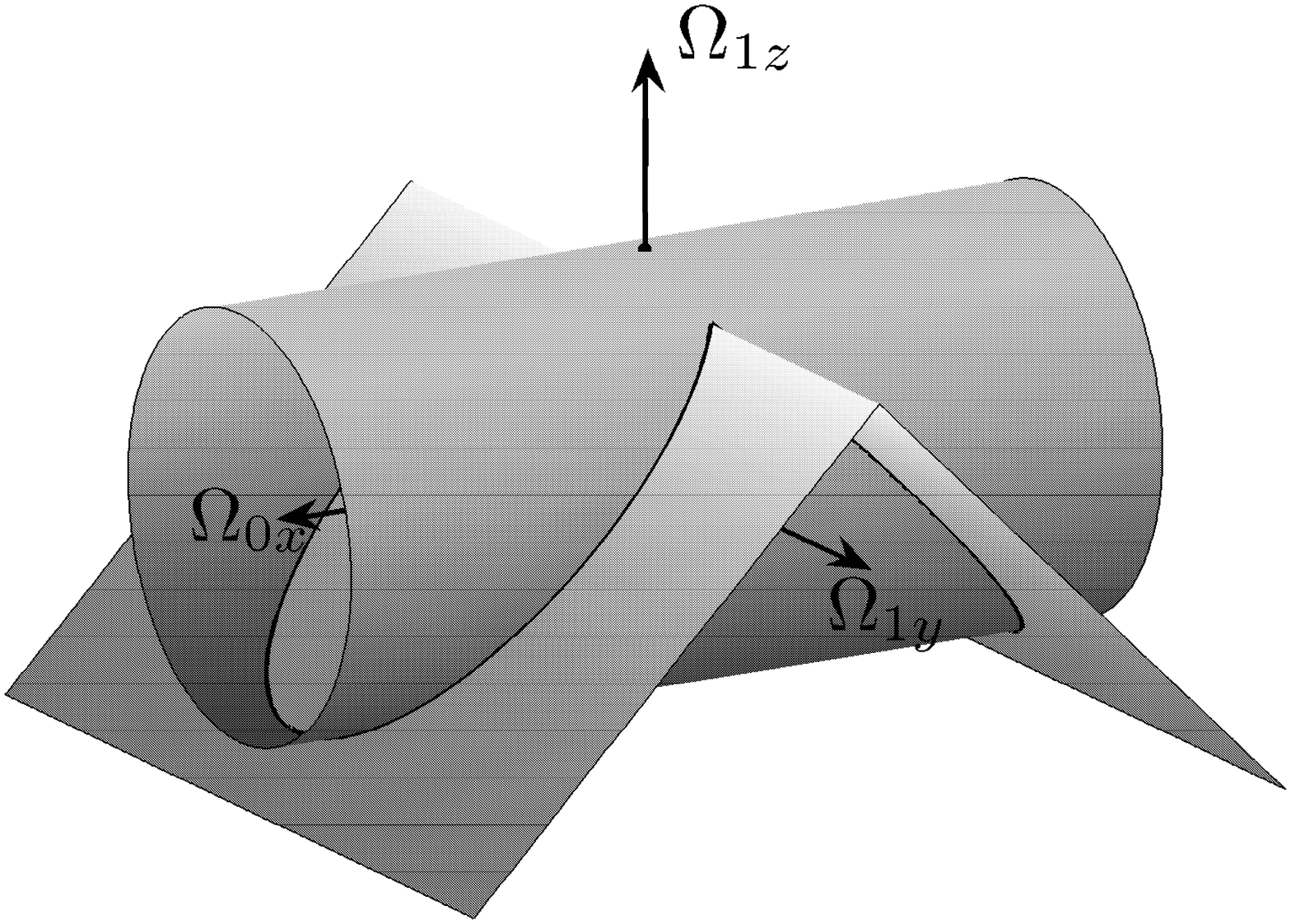}
\includegraphics[angle=-90,scale=0.3]{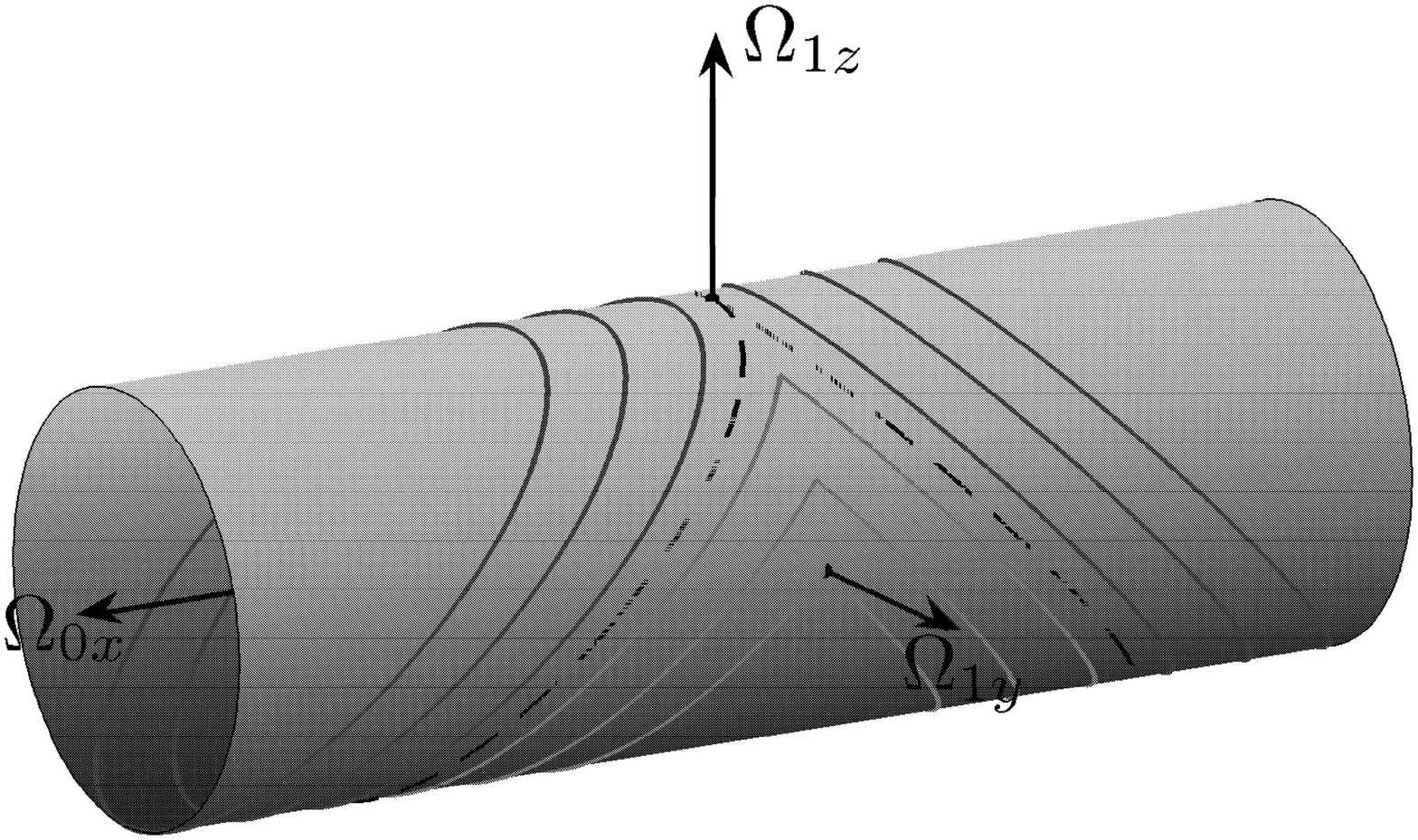}
\caption{Intersection of the two surfaces associated with the first integrals (top), and phase portrait of the system~\eqref{systTO11f} (bottom) plotted on the cylinder of radius $1$.\label{figA5}}
\end{figure}

The robust time-optimal solution at first order is thus a pulse along the axis $\vec{e}_x$ of constant amplitude. Some switches between the maximum and the minimum values of the field can exist. In this case, the differential system becomes:
\begin{equation}
\dot{\Omega}_{0x}=\Omega_{1y},~\dot{\Omega}_{1y}=\sign(\Omega_{0x})\Omega_{1z},~\dot{\Omega}_{1z}=-\sign(\Omega_{0x})\Omega_{1y},
\label{systTO11f}
\end{equation}
and the first integrals are given by:
\begin{equation}
H=|\Omega_{0x}|+\Omega_{1z},\quad 1=\Omega_{1y}^2+\Omega_{1z}^2.
\end{equation}
Note that the conservation of $H$ is associated with two semi-planes intersecting along the line of equation $(\Omega_{0x},\Omega_{0z})=(0,H)$, as shown in Fig.~\ref{figA5}.

\begin{figure}[h!]
\centering
\includegraphics[angle=-90,scale=0.4]{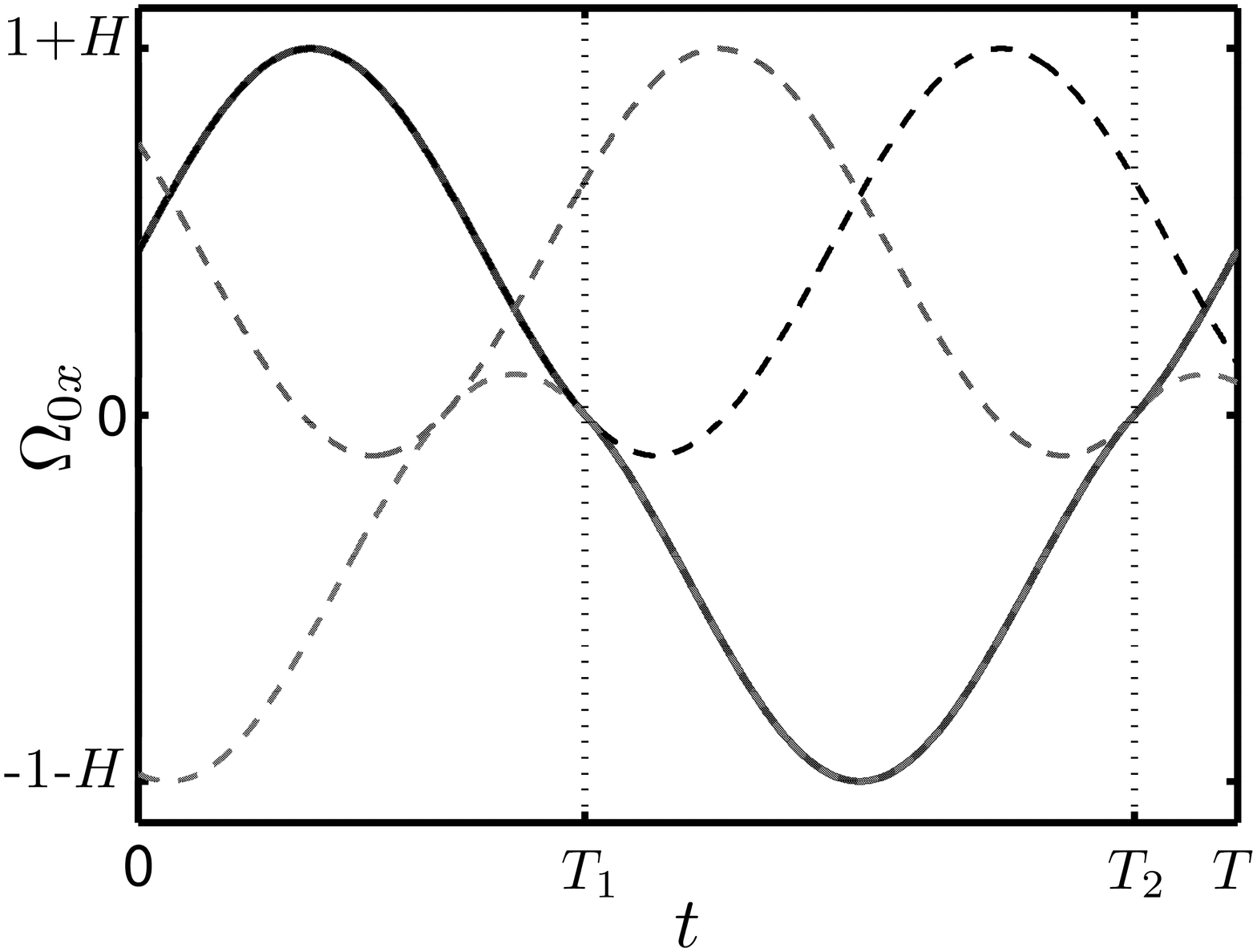}
\caption{Solution $\Omega_{0x}(t)$ over one period (solid blue line). The curves of equations $(\Omega_{0x}=H+\sin t)$, $(\Omega_{0x}=\sin(t-2T_1)-H)$ and $(\Omega_{0x}=H+\sin(t+2(T_1-T_2)))$ are respectively plotted in dashed black line, dashed gray line and dashed light gray line.\label{figA6}}
\end{figure}

If $H>1$, the sign of $\Omega_{0x}$ remains the same for any time $t$ which means that no switch appears. The control field $u_x$ is a constant pulse and it cannot realize a robust inversion. We now analyze the case $H\in[0,1]$, and we explicitly derive a solution starting at time $t=0$ from the point $\Omega_{0x}(0)=H$ and $\Omega_{0y}(0)=1$. We consider one period, that is one cycle of a trajectory. For one period, we can see that the sign of $\Omega_{0x}$ changes two times. We denote by $T_1$ and $T_2$ the times of the first and second switches. The solutions can be expressed as follows:
\begin{equation}
\begin{cases}
\Omega_{0x}=H+\sin t,\\
\Omega_{1y}=\cos t,\\
\Omega_{1z}=-\sin t,
\end{cases}
\end{equation}
for $t\in[0,T_1]$,
\begin{equation}
\begin{cases}
\Omega_{0x}=\sin(t-2T_1)-H,\\
\Omega_{1y}=\cos(t-2T_1),\\
\Omega_{1z}=\sin(t-2T_1),
\end{cases}
\end{equation}
for $t\in[T_1,T_2]$ and
\begin{equation}
\begin{cases}
\Omega_{0x}=H+\sin(t+2(T_1-T_2)),\\
\Omega_{1y}=\cos(t+2(T_1-T_2)),\\
\Omega_{1z}=-\sin(t+2(T_1-T_2)),
\end{cases}
\end{equation}
for $t\in[T_2,T]$. Note that the solutions of the system~\eqref{systTO11f} are given for any initial condition so that $\Omega_{0x}(0)=H$ and $\Omega_{0y}(0)=1$. The times $T_1$ and $T_2$ correspond to the switches of $\Omega_{0x}$. $T$ is the period of the solution. We have $T_1=\pi+\arctan\left(\tfrac{H}{\sqrt{1-H^2}}\right)$, $T_2=3T_1-\pi$ and $T=4T_1-2\pi$. The evolution of $\Omega_{0x}(t)$ is displayed in Fig.~\ref{figA6}.
To summarize, the control field is a pulse of constant amplitude with switches at times $T_1$ and $T_2$ given in terms of $H$.
The final goal is then to determine the value of $H$ which allows us to realize a robust transfer in minimum time. We start by integrating the dynamics of the states $\vec{q}_0(t)$ and $\vec{q}_1(t)$ in the system~\eqref{eq1} with a control field given by $u_x=\sign(\Omega_{0x})$. The system is of the form:
\begin{equation}
\dot{y}_0=\sign(\Omega_{0x})z_0,~\dot{z}_0=-\sign(\Omega_{0x})y_0,~\dot{x}_1=y_0.
\end{equation}
Introducing the angle $\theta=\arctan(y_0/z_0)$, we can show that the system becomes:
\begin{equation}
\dot{\theta}=\sign(\Omega_{0x}),~\dot{x}_1=\sin\theta.
\label{eqstateTO1}
\end{equation}
In these coordinates, a robust transfer is of the form $\theta(0)=0\rightarrow\theta(t_f)=\pi$ and $x_1(0)=0\rightarrow x_1(t_f)=0$. The solutions of this equation are:
\begin{equation}
\begin{cases}
\theta=t,\\
x_1=1-\cos(t),\\
\end{cases}
\end{equation}
for $t\in[0,T_1]$,
\begin{equation}
\begin{cases}
\theta=-t+2T_1,\\
x_1=1+2\sqrt{1-H^2}+\cos(t-2T_1),\\
\end{cases}
\end{equation}
for $t\in[T_1,T_2]$ and
\begin{equation}
\begin{cases}
\theta=t+2(T_1-T_2),\\
x_1= 1+4\sqrt{1-H^2}-\cos(t+2(T_1-T_2)),\\
\end{cases}
\end{equation}
for $t\in [T_2,T]$.

\begin{figure}[h!]
\centering
\includegraphics[angle=-90,scale=0.4]{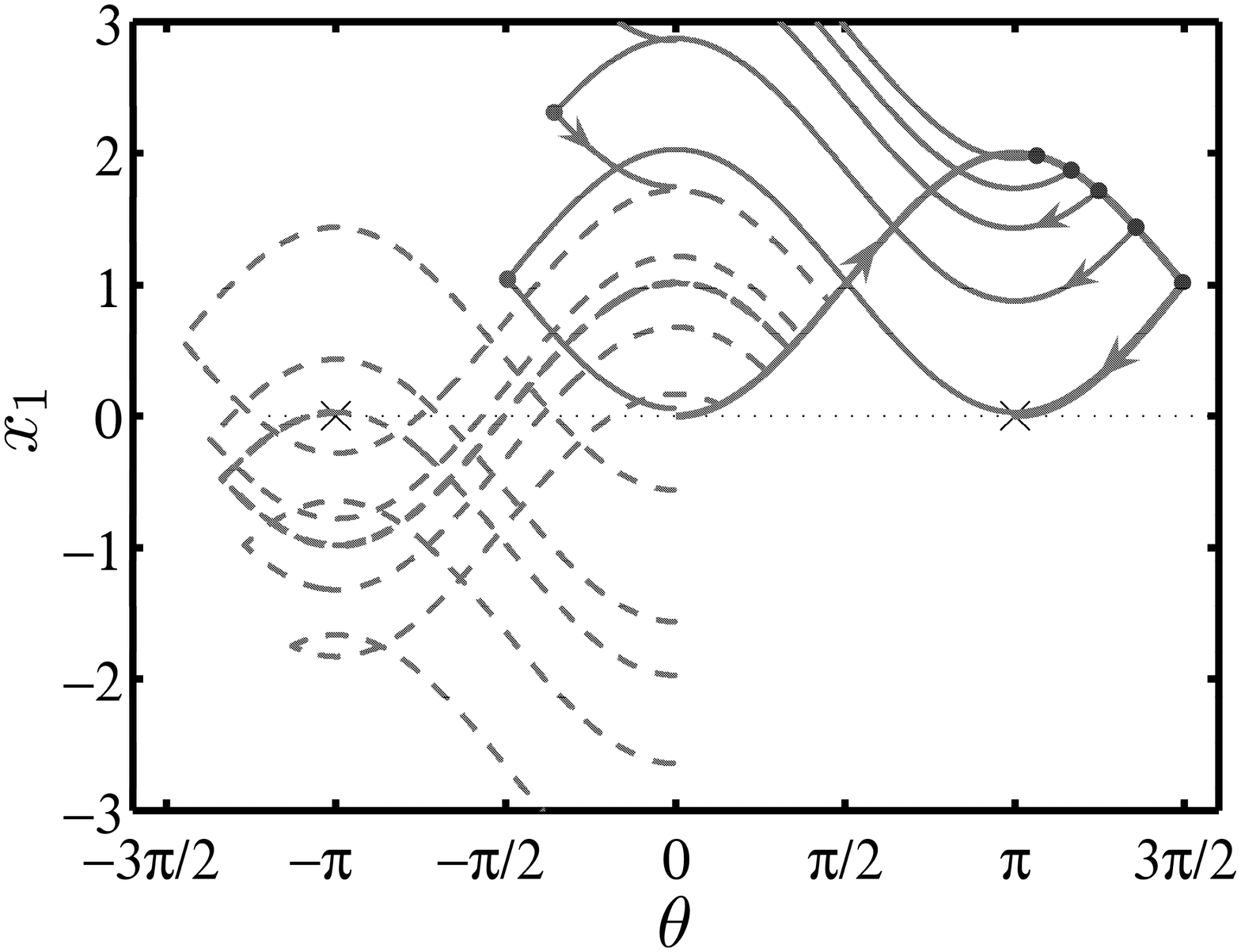}
\caption{Solutions of Eq.~\eqref{eqstateTO1} in the plane $(\theta,x_1)$. The crosses represent the target states. Every trajectories start at the origin $(0,0)$ of this plane. The solid lines are the solutions described in the text. The dashed lines correspond to the trajectories associated to some solutions of the system~\eqref{systTO11f} starting on a point $\Omega_{0x}(0)=H$ and $\Omega_{1y}(0)=-1$. The dark gray point corresponds to the first bang, and the  light gray point to the second one. The golden/brown thick solid line is the global optimal solution, and the dashed one a sub-optimal solution.\label{figA7}}
\end{figure}

We plot in Fig.~\ref{figA7} the solutions in the plane $(\theta,x_1)$ for different values of $H$ in order to find the optimal solution.
We obtain that the global optimum corresponds to a very particular case, which occurs in the limit $H\to 1$. The corresponding control field is associated with a trajectory infinitely close to the separatrix of Fig.~\ref{figA5}. Note that $H$ cannot be exactly equal to $1$, because, in this case, there is no switch. The solution satisfies $T_1\to 3\pi/2$ and $t^*\to 2\pi$. This point is shown in Fig. 3 of the main text. The duration $t^*=2\pi$ is thus the physical minimum time to make a first-order robust inversion with a control field bounded by $1$. The computation of the second and third order robust parameters is made with a numerical gradient algorithm.
\section{Robustness against control field inhomogeneities\label{Sect-inhom}}\label{appc}
In this paragraph, we derive the robust optimal field at first order. The differential system to solve is given by:
\begin{equation}
\begin{cases}
\dot{\Omega}_{0x}=-\Omega_{0y}(\Omega_{0z}+\Omega_{1z}),\\
\dot{\Omega}_{0y}=\Omega_{0x}(\Omega_{0z}+\Omega_{1z}),\\
\dot{\Omega}_{0z}=\Omega_{0y}\Omega_{1x}-\Omega_{0x}\Omega_{1y},\\
\dot{\Omega}_{1x}=-\Omega_{0y}\Omega_{1z},\\
\dot{\Omega}_{1y}=\Omega_{0x}\Omega_{1z},\\
\dot{\Omega}_{1z}=\Omega_{0y}\Omega_{1x}-\Omega_{0x}\Omega_{1y}.
\end{cases}
\end{equation}
Note that the control landscape is parameterized by the two initial values of $\Omega_{1x}(0)$ and $\Omega_{1y}(0)$.
In addition to the Pontryagin's Hamiltonian, this system has $5$ first integrals which are $\Omega_{0z}-\Omega_{1z}=0$, $I_x=\Omega_{0x}-2\Omega_{1x}$, $I_y=\Omega_{0y}-2\Omega_{1y}$, $J=\vec{\Omega}_0\cdot\vec{\Omega}_1$ and $M=|\vec{\Omega}_1|$. The relation $\Omega_{0z}-\Omega_{1z}=0$ is due to the initial conditions $\Omega_{kz}(0)=0$ for all $k$. Using these constants, the system becomes:
\begin{equation}
\begin{cases}
\dot{\Omega}_{0x}=-2\Omega_{0y}\Omega_{0z},\\
\dot{\Omega}_{0y}=2\Omega_{0x}\Omega_{0z},\\
\dot{\Omega}_{0z}=-\tfrac{1}{2}I_x\Omega_{0y}+\tfrac{1}{2}I_y\Omega_{0x}.
\end{cases}
\end{equation}
Since $\Omega_{0x}(0)=1$ and $\Omega_{0y}(0)=\Omega_{0z}(0)=0$, a solution depends only on the two parameters $I_x$ and $I_y$ and we can show that $2J=1-I_x$. This system can be integrated in terms of Jacobi's elliptic functions. We introduce the parameters $\omega$ and $m$ so that:
\begin{equation}
\omega=(I_x^2+I_y^2)^{\frac{1}{4}},\quad m=\tfrac{1}{2}-\tfrac{I_x}{2\omega^2},
\end{equation}
and the Jacobi's amplitude function:
\begin{equation}
\nu=\am(\omega t+\K(m),m).
\end{equation}
We also introduce the phase of the control field $\phi$ so that $\Omega_{0x}=\cos\phi$ and $\Omega_{0y}=\sin\phi$.
It can be checked that the following functions are solutions of the problem:
\begin{eqnarray}
& & \phi=-2\left[\tfrac{\sin\nu}{|\sin\nu|}\arccos(\sqrt{1-m\sin^2\nu})-\arccos(\sqrt{1-m})\right],\nonumber\\
& & \Omega_{0z}=-\omega\sqrt{m}\cos\nu.
\label{fieldO1alpha}
\end{eqnarray}
At time $t=t_f$, we must have $\Omega_{0z}(t_f)=0$ since the Bloch vector reaches the south pole of the Bloch sphere. Furthermore, $\Omega_{0z}$ goes back to zero when the time $t$ is a multiple of $2\K(m)/\omega$.

In order to locate the global optimum, we use a similar method as in Sec.~\ref{sect-O2E}. We integrate the system~\eqref{edqialpha} with the control fields~\eqref{fieldO1alpha} and we compute the corresponding fidelity for every pair $(I_x,I_y)$. We obtain that the optimal control field verifies:
\begin{equation}
I_x=0.6995,\quad I_y=1.1192,\quad t_f=\tfrac{4\K(m)}{\omega}=1.86\pi.
\end{equation}
The result is presented in the left panels of Fig. 4. We use a numerical algorithm to compute the solutions robust at second and third order. 

\end{document}